\documentclass[acmlarge]{acmart}
\usepackage{algorithm}
\usepackage{url}
\usepackage{longtable}
\usepackage{enumerate}
\usepackage{enumitem}
\usepackage{array}
\usepackage{algpseudocode}
\usepackage{algorithmicx}
\usepackage{multirow}
\usepackage{array}
\usepackage{soul,color}
\usepackage{subfigure}
\usepackage{overpic}
\usepackage{bm}
\usepackage{graphicx}
\usepackage{booktabs}
\usepackage{threeparttable}
\usepackage{wrapfig}
\usepackage{stfloats}
\usepackage{amsthm}

\newtheorem{definition}{Definition}
\allowdisplaybreaks[4]
\def\BibTeX{{\rm B\kern-.05em{\sc i\kern-.025em b}\kern-.08em
    T\kern-.1667em\lower.7ex\hbox{E}\kern-.125emX}}
\usepackage{balance}

\AtBeginDocument{%
  \providecommand\BibTeX{{%
    Bib\TeX}}}

\begin{document}

\title{A Comprehensive Survey on Edge Data Integrity Verification: Fundamentals and Future Trends}

\author{Yao Zhao}
\affiliation{%
  \institution{School of Information Technology, Deakin University}
  \country{Australia}}

\author{Youyang Qu}
\affiliation{
  \institution{1. Key Laboratory of Computing Power Network and Information Security, Ministry of Education, Shandong Computer Science Center, Qilu University of Technology (Shandong Academy of Sciences), 2. Shandong Provincial Key Laboratory of Computer Networks, Shandong Fundamental Research Center for Computer Science}
  \country{China}
}

\author{Yong Xiang}
\affiliation{%
 \institution{School of Information Technology, Deakin University}
 \country{Australia}}

\author{Md Palash Uddin}
\affiliation{%
 \institution{School of Information Technology, Deakin University}
 \country{Australia}}

\author{Dezhong Peng}
\affiliation{%
 \institution{College of Computer Science, Sichuan University}
 \country{China}}

\author{Longxiang Gao}
\authornote{Longxiang Gao and Youyang Qu are the co-corresponding authors.}
\affiliation{%
  \institution{1. Key Laboratory of Computing Power Network and Information Security, Ministry of Education, Shandong Computer Science Center, Qilu University of Technology (Shandong Academy of Sciences), 2. Shandong Provincial Key Laboratory of Computer Networks, Shandong Fundamental Research Center for Computer Science}
  \country{China}}

\begin{abstract}
Recent advances in edge computing~(EC) have pushed cloud-based data caching services to edge, however, such emerging edge storage comes with numerous challenging and unique security issues. One of them is the problem of edge data integrity verification (EDIV) which coordinates multiple participants (e.g., data owners and edge nodes) to inspect whether data cached on edge is authentic. To date, various solutions have been proposed to address the EDIV problem, while there is no systematic review. Thus, we offer a comprehensive survey for the first time, aiming to show current research status, open problems, and potentially promising insights for readers to further investigate this under-explored field. Specifically, we begin by stating the significance of the EDIV problem, the integrity verification difference between data cached on cloud and edge, and three typical system models with corresponding inspection processes. To thoroughly assess prior research efforts, we synthesize a universal criteria framework that an effective verification approach should satisfy. On top of it, a schematic development timeline is developed to reveal the research advance on EDIV in a sequential manner, followed by a detailed review of the existing EDIV solutions. Finally, we highlight intriguing research challenges and possible directions for future work, along with a discussion on how forthcoming technology, e.g., machine learning and context-aware security, can augment security in EC. Given our findings, some major observations are: there is a noticeable trend to equip EDIV solutions with various functions and diversify study scenarios; completing EDIV within two types of participants (i.e., data owner and edge nodes) is garnering escalating interest among researchers; although the majority of existing methods rely on cryptography, emerging technology is being explored to handle the EDIV problem.
\end{abstract}



\maketitle

\section{Introduction}
\begin{figure}[!tp]
    \centering
    \includegraphics[width=0.7\linewidth]{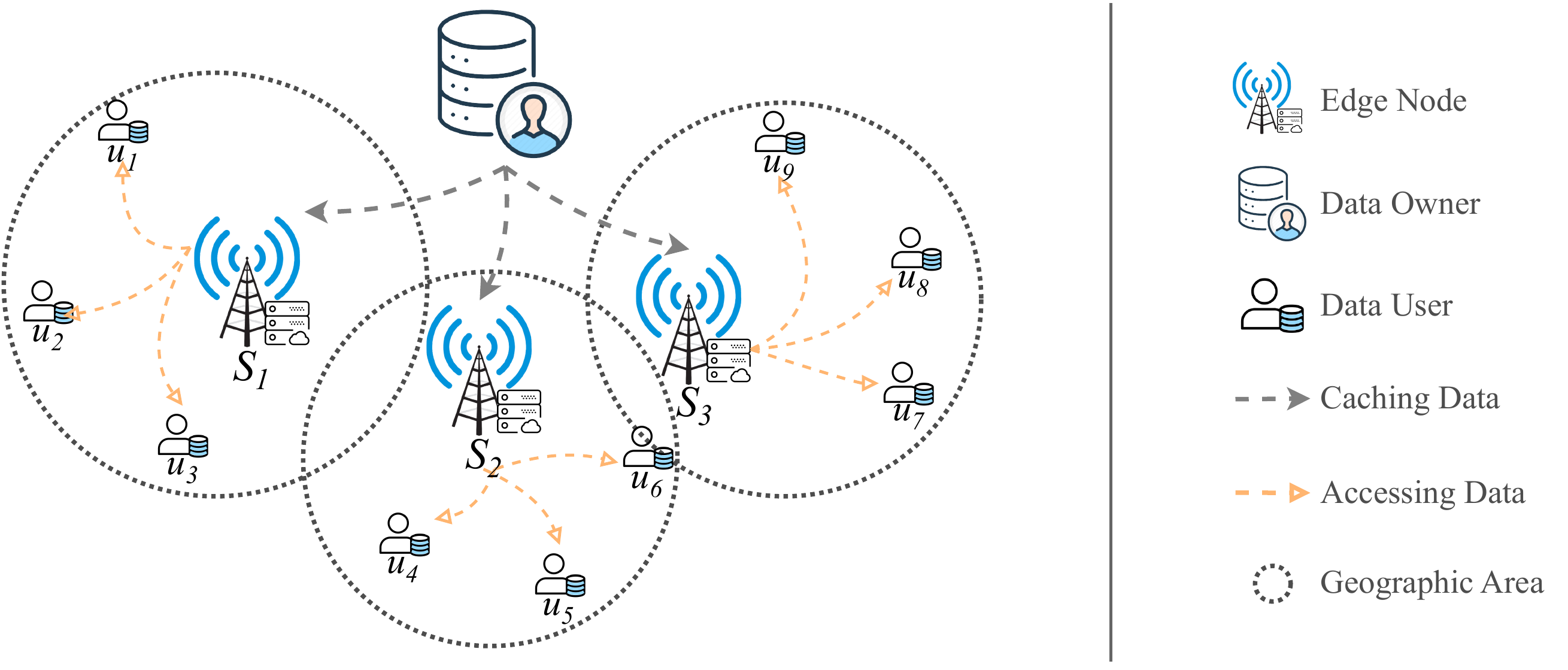}
    \caption{Example of edge storage. A data owner caches multiple data replicas to geographically distributed edge nodes (denoted by $S_1$, $S_2$, $S_3$) to serve nearby data users (denoted by $u_i,\ i\in \{1,2,\cdots,9\}$) with ultra-low data access latency.}
    \label{fig:System model}
\end{figure}

The global deployment of mobile and \textbf{Internet-of-Things (IoTs)} devices has been rapidly increasing as a result of the current expansion of 5G and beyond networks~\cite{kong2022edge}. According to the technical report~\cite{laghari2021review}, the number of these devices is projected to surpass 25.4 billion by 2030. These devices serve as fundamental components for smart applications to carry out the most basic yet essential activities such as detecting~\cite{guo2020detecting}, actuating~\cite{oliveira2015iot}, and controlling~\cite{verma2019sensing}. However, it is insufficient to depend just on those low-performance devices to complete complex activities efficiently, e.g., smart transportation arrangements~\cite{zhang2020architecture, aamir2019sustainable, babar2019real}, smart medical treatments~\cite{onasanya2021smart, jahan2021collaborative, gai2019toward}, and smart vehicle control~\cite{raza2019survey, baidya2020vehicular, lu2022analytical}. High-performance computing infrastructures are essential for offloading calculation tasks and facilitating decision-making. Undoubtedly, \textbf{cloud computing (CC)}~\cite{zou2021integrated, toosi2014interconnected, kabir2021uncertainty} is the most well-known of such technologies. In CC environments, \textbf{cloud infrastructure providers (CIPs)}, e.g., One Drive\footnote{https://www.microsoft.com/en-us/microsoft-365/onedrive/online-cloud-storage}, Amazon\footnote{https://aws.amazon.com/}, and Google Drive\footnote{https://www.google.com/drive/}, deliver centralized data caching services to support large-scale data access~\cite{haber2022cloud, boss2007cloud}. Yet, cloud computing is incapable of perfectly matching the demands of mobile/IoT services due to the concerns like geographical unawareness~\cite{kaur2020energy}, bandwidth limitations~\cite{qi2012research}, a lack of real-time services~\cite{tahirkheli2021survey}, and unpredictable data access latency~\cite{choo2010cloud}. To this end, an emerging paradigm named \textbf{edge computing (EC)}~\cite{cao2020overview, hua2022edge, cruz2022edge} is spawned as one of the key enabler technologies for 5G and beyond to facilitate latency-sensitive or geo-aware applications, e.g., autopilot~\cite{wang2019ecass}, virtual reality~\cite{martin2022multimodality}, and video analytics~\cite{ananthanarayanan2017real}. The detailed definition and origin of EC can refer to~\cite{satyanarayanan2017emergence}. Motivated by EC, \textbf{data owners (DOs)} are allowed to outsource popular data on \textbf{edge nodes (ENs)} for serving nearby \textbf{data users (DUs)} with better user experience~\cite{hassan2019edge}, as shown in Fig.~\ref{fig:System model}. Due to such benefits over CC, EC has grown dramatically in recent years~\cite{siriwardhana2021survey}. The Market Study Report\footnote{https://www.gminsights.com/industry-analysis/edge-data-center-market} predicts that the edge data centre market is expected to exceed \$20 billion by 2026.

\par Unfortunately, this promising computing paradigm still faces alarming security challenges in practice~\cite{zeyu2020survey, xiao2019edge, husain2021survey, mukherjee2020intelligent}. Different from cloud facilitated by mega-scale data centres, edge nodes are usually deployed at base stations or access points and managed by different \textbf{edge infrastructure providers (EIPs)}~\cite{cao2020edge}. This edge caching strategy is much more distributed, dynamic, and volatile~\cite{liu2019survey}, making the integrity of cached data corrupted easily and frequently. Furthermore, various attacks against EC-related infrastructures have significantly increased recently~\cite{qiu2020edge}. For instance, Mirai virus~\cite{antonakakis2017understanding, stiles2019hardware, al2019botnet}, released in August 2016 and managed to infiltrate more than 65,000 IoT devices within the first 20 hours of that release, is one of the most famous assaults to have ever taken place in reality. Over 178,000 domains were knocked down as a result of DDoS assaults launched against edge nodes a few days later using botnets created from these infected devices~\cite{chinese2019development}. IoTReaper and Hajime, two Mirai variants that were discovered shortly after, were thought to have infected more than 378 million IoT devices in 2017~\cite{xiao2019edge, mudgerikar2021iot}. These IoT botnet assaults were estimated to have cost over 100 million USD in damages since the initial Mirai botnet was found in 2016~\cite{anderson2019measuring, tuptuk2018security, xiao2019edge}. More intuitively, researchers have found that various factors may lead to data loss in real-world scenarios. Based on the report from Kroll Ontrack\footnote{https://www.techradar.com/how-to/world-of-tech/management/how-to-recover-lost-business-data-1304303/2}, 67\% data loss is attributed to hard drive crashes or system failure, 14\% is blamed for human error, and 10\% is a result of software failure.

\par The aforementioned examples and statistics clearly illustrate the unsatisfactory state of edge data security. Outsourcing data to edge nodes results in the separation of data ownership and management. Thus, data owners and data users may not always trust edge nodes, since they may misuse data management permissions and expose data to security risks~\cite{zhang2021resource}. Consequently, a variety of issues must be addressed before subscribing to edge data caching services. For example, \emph{how do data owners trust EIPs and ensure that outsourced data is integral all the time?} \emph{How to properly audit cached data without retrieving the whole data collection?} \emph{How to maintain the stable operation of integrity audit while data owners modify outsourced data?} All these challenges could be summarized to the \textbf{edge data integrity verification (EDIV)} problem~\cite{li2020auditing}, which is defined as follows.
\begin{definition}[Edge Data Integrity Verification Problem]
The edge data integrity verification problem refers to inspecting the accuracy and consistency~(validity) of data replicas cached on edge nodes.
\end{definition}

\par An EDIV approach entails creating a solution that allows data owners or (and) users to verify the integrity of outsourced data within the edge environment. EDIV investigation is of particularly practical importance for edge-based services/applications, because critical business decisions depend mostly on accurate edge data. If the data integrity is compromised, any decision based on that becomes questionable. We further emphasize its significance in Section~\ref{subsec:The Significant of Edge Data Integrity}. To date, numerous great achievements have been made for the EDIV problem, such as verification efficiency improvement~\cite{li2020auditing} and data privacy guarantee~\cite{tong2022privacy}, however, all these articles have proposed specific EDIV solutions tailored to their respective domains, exposing the lack of a systematic and comprehensive review of them. 

\subsection{Related Surveys and Our Contributions}
\label{subsec:Contributions and Paper Organization}
In the context of data integrity verification, the \textbf{cloud data integrity verification (CDIV)} problem has received significant attention over the past decade, resulting in several related surveys~\cite{zhou2018data, pujar2020survey, kumar2020extensive, zafar2017survey, gudeme2019review, li2020survey, kumar2020systematic, gangadevi2021survey, ghallab2021data, han2022survey}. For example, Suchetha \emph{et al.}~\cite{pujar2020survey} offered an overview of various techniques for data integrity verification in cloud storage. Gangadevi \emph{et al.}~\cite{gangadevi2021survey} conducted a brief survey on data integrity verification schemes for cloud computing based on blockchain technology. Han \emph{et al.}~\cite{han2022survey} provided a survey of blockchain-based integrity auditing for cloud data recently, including evaluation criteria, a review of existing solutions, and suggestions for future research directions. Despite the comprehensive surveys on CDIV, EDIV exhibits fundamental differences from CDIV, as articulated in Section~\ref{subsec:Edge Data Integrity Versus Cloud Data Integrity}.

\par Motivated by it, this work aims to provide new insights into data integrity in edge computing domains. To our best knowledge, this is the first survey on edge data integrity verification. In this survey, we begin with describing the motivation for studying EDIV problems and providing a comprehensive comparison between CDIV and EDIV. Afterward, three typical system models with corresponding key processes are covered, together with a set of criteria that an effective EDIV approach should satisfy. After that, depending on the design objectives, we comb through a taxonomy of EDIV solutions, ranging from 2019 to 2023. Finally, we highlight unresolved challenges and make recommendations for further research. This is done to clarify the link between CDIV and EDIV, as well as to promote future development and integration of EDIV. More significantly, we offer a valuable resource for follow-up researchers and amateurs. The primary contributions of this survey are overviewed as follows.
\begin{itemize}
  \item
    We clarify the gravity and significance of the EDIV study and summarize its uniqueness compared with CDIV. In addition, the system models along with the key processes of EDIV are introduced in detail.
  \item
    We synthesize a comprehensive criteria system that a satisfactory EDIV solution is expected to meet, which can be further applied to assess the effectiveness of EDIV methods.
  \item
    According to the established criteria, a chronological timeline is given to outline the evolution of existing endeavors for the EDIV problem. Furthermore, we systematically categorize current solutions into three types, emphasizing their strengths while exposing their shortcomings.
  \item
    We identify a list of open issues and further exploit future research directions including traditional and outspread ones to promote dedicated efforts on the EDIV problem. Notably, some of the valuable directions have barely or even never been investigated yet. We hope it could provide insights for follow-up researchers.
\end{itemize}

\begin{figure}[!tp]
    \centering
    \includegraphics[width=1\linewidth]{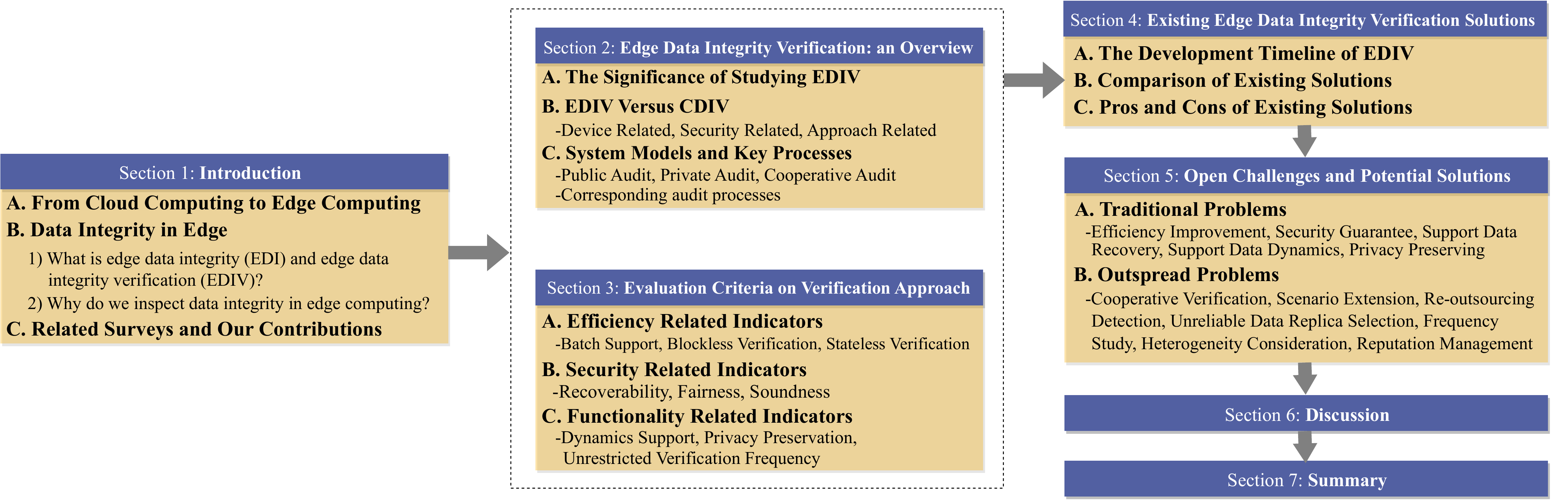}
    \caption{Roadmap of the survey}
    \label{fig:Roadmap of the survey}
    \vspace{-1 em}
\end{figure}

\begin{table}[!tp]
\vspace{-1 em}
\small
\renewcommand\arraystretch{1.2}
\caption{List of key abbreviations}
\label{tab:List of key abbreviations}
\setlength{\tabcolsep}{1mm}{
\begin{tabular}{cl|cl|cl}
\hline\hline
\textbf{Abbr.} & \multicolumn{1}{c|}{\textbf{Definition}} & \textbf{Abbr.} & \multicolumn{1}{c|}{\textbf{Definition}} & \textbf{Abbr.} & \multicolumn{1}{c}{\textbf{Definition}} \\ \hline
IoTs            & Internet-of-Things                       & CC             & Cloud Computing                          & CIP            & Cloud Infrastructure Provider           \\ \hline
EC             & Edge Computing                           & DO             & Data Owner            & EN             & Edge Node                     \\ \hline
DU             & Data User                  & EIP            & Edge Infrastructure Provider                       & EDIV            & Edge Data Integrity Verification              \\ \hline
CDIV            & Cloud Data Integrity Verification               & SLA            & Service Level Agreement                          & TPA            & Third Party Auditor                           \\ \hline
EDI            & Edge Data Integrity                                    & PDP            & Provable Data Possession                 & POR            & Proof of Retrievability                 \\\hline\hline
\end{tabular}}
\end{table}

\subsection{Paper Organization}
\label{subsec:Paper Organization}
The remainder of this survey is structured as follows. The motivation and overview of the EDIV problem are presented in Section~\ref{sec:Edge data integrity: an Overview}. In Section~\ref{sec:Evaluation Criteria on Verification Approaches}, we propose a series of criteria regarding the evaluation of existing EDIV solutions. A development timeline and taxonomy on EDIV are summarized and the existing works are reviewed accordingly in Section~\ref{sec:Existing EDIV solutions}. In Section~\ref{sec:Open challenges and potential solutions}, several future research directions and potential solutions are introduced, while the impact of emerging technologies in enhancing security for EC is discussed in Section~\ref{sec:Discussion}. Finally, a summary is provided in Section~\ref{sec:Conclusion}. For clarity, we illustrate the organization of this work in Fig.~\ref{fig:Roadmap of the survey}, and key acronyms are outlined in Table~\ref{tab:List of key abbreviations}.

\section{Edge data integrity Verification: an Overview}
\label{sec:Edge data integrity: an Overview}
To better understand the scope and breadth of the EDIV problem, in this section, we state the significance of EDIV investigation. Then, we explicitly present the discrepancy between CDIV and EDIV. Further, we provide a summary of three commonly-used system models, along with a short introduction to the corresponding key processes concerning the EDIV problem-solving strategies.

\subsection{The Significance of Edge Data Integrity Verification}
\label{subsec:The Significant of Edge Data Integrity}
To some extent, data cached on cloud is more reliable and stable than on edge nodes~\cite{xia2019secure, huh2019understanding}, since cloud servers have adequate resources to achieve computation-intensive inspection tasks, while edge nodes often can not afford to perform the same level of integrity assurance~\cite{ni2020security}. In reality, however, data corruption accidents occur frequently even in cloud. According to a comprehensive study~\cite{wang2015understanding}, existing cloud data corruption detection schemes are quite insufficient. Specifically, only 25\% of data corruption problems are reported correctly, 42\% are undetected, and 21\% receive imprecise error reports. They also found that the detection system raises 12\% false alarms. Real examples include but not limited to the following ones. Jeff Bonwick, the ZFS\footnote{https://en.wikipedia.org/wiki/ZFS} creator, mentioned that a fast database named Greenplum\footnote{https://greenplum.org/} faces undetected data corruption every 10 to 20 minutes\footnote{https://queue.acm.org/detail.cfm?id=1317400}. Additionally, NetApp\footnote{https://www.netapp.com/} conducted 41-month real-world research on more than 1.5 million hard disk drives and identified over 400,000 undiscovered data corruptions, including more than 30,000 undetectable by hardware RAID controllers~\cite{bairavasundaram2008analysis}. Besides, during the course of six months and involving around 97 petabytes of data, CERN\footnote{http://home.cern/} discovered that approximately 128 megabytes of data got irreversibly corrupted~\cite{bairavasundaram2007analysis}.

\par The above analysis clearly reveals that detecting data corruption is a challenging problem in cloud domains, let alone in dynamic edge computing environments. Briefly, studying the EDIV problem has the following significance from the utility perspective.

\par \emph{Cut Data Owners' Loss}. Edge data corruption has a lasting impact on data owners' businesses. For recoverable data, detecting corruption as soon as possible can help data owners recover correct data in a timely way so that effective measures can be taken to shrink the gaps left by corruption~\cite{xu2017nova, bairavasundaram2008analysis}. For unrecoverable data, identifying corruption efficiently can assist data owners in designing emergency plans to minimize unnecessary delay and possible loss of business reputation and revenue~\cite{pankowska1999outsourcing}. Furthermore, inspecting \textbf{edge data integrity (EDI)} presents a critical aspect to reducing the customer churn rate, increasing data users' trust and respect, and reestablishing client relationships.

\par \emph{Boost EIPs' Reputation}. In practical terms, there are thousands of EIPs around the world, each of which is renowned in certain geographic areas. For instance, Optus~\cite{hutchinson1994telecommunications} is highly accepted in Australia, and IBM is more prestigious in America. Business competition can be fierce, especially in fast-moving edge computing markets where data owners often shop around for cost-effective EIPs~\cite{cui2022application}. A satisfying EDIV solution can help EIPs defend their market position and build their competitive advantage.

\par \emph{Remedy Deficiency}. In real production environments, edge nodes adopt internal data and metadata checksumming~\cite{sivathanu2004enhancing} to detect data corruption~\cite{ghemawat2003google}. In some cases, although EIPs have bounded by \textbf{service level agreements (SLAs)}~\cite{bianco2008service, nugraha2021towards} to ensure data integrity, data owners can not solely rely on such agreements, because edge data operational details are not transparent to the data owners and EIPs may be untrusted~\cite{ma2018privacy}, i.e., EIPs may not conduct required data integrity check mechanisms to actively report corruption for keeping a good industry reputation. Even if EIPs are assumed to be totally honest and self-actualized, outsourced data replicas could be manipulated or lost due to accidental activities~\cite{yuan2021coopedge}, which can be a nightmare for data owners and an embarrassment for EIPs. Thus, an effective EDI external verification approach can be regarded as a supplement to internal checksumming, supporting data owners and EIPs to detect corruption quickly.

\begin{table}[]
\small
\centering
\renewcommand\arraystretch{1.2}
\caption{Cloud Data Integrity Verification Versus Edge Data Integrity Verification}
\label{tab:Cloud Data Integrity vs Edge Data Integrity}
\setlength{\tabcolsep}{1mm}{
\begin{tabular}{cc|c|l|l}
\hline \hline
\multicolumn{2}{c|}{\textbf{Cat.}}                                     & \multicolumn{1}{c|}{\textbf{Sub-category}} & \multicolumn{1}{c|}{\textbf{CDIV}} & \multicolumn{1}{c}{\textbf{EDIV}} \\ \hline
\multirow{6}{*}{\textbf{\rotatebox{90}{Device}}}   & \multirow{6}{*}{\textbf{\rotatebox{90}{Related}}} & The number of devices   & $\bullet$ limited number of cloud servers & $\bullet$ large-scale edge nodes                 \\ \cline{3-5} 
 &     & Device location                            & \begin{tabular}[c]{@{}l@{}}$\bullet$ remote and centralized, long distance\\ \ \ \ from data users, less mobility\end{tabular}              & \begin{tabular}[c]{@{}l@{}}$\bullet$ at the edge of the network and distributed,\\ \ \ \ close to data users, high mobility\end{tabular}                                                \\ \cline{3-5} 
                                   &   & Device capacity                            & \begin{tabular}[c]{@{}l@{}}$\bullet$ more secure, less scalability, more\\ \ \ \ latency, virtually unlimited resources\end{tabular} 
    & \begin{tabular}[c]{@{}l@{}}$\bullet$ less secure, more scalability, less latency,\\ \ \ \ fewer resources  \end{tabular}                                                 \\ \cline{3-5} 
                                   &    & Device provider                           & $\bullet$ big companies                                                                                                        & $\bullet$ less powerful entities, e.g., small companies                                \\ \hline
\multirow{4}{*}{\textbf{\rotatebox{90}{Security}}} & \multirow{4}{*}{\textbf{\rotatebox{90}{Related}}} & The number of risks                        & $\bullet$ less outside security risks                                                                                          & $\bullet$ more risks, more likely to be attacked  \\ \cline{3-5} 
                                   &                                  & The type of risks               & \begin{tabular}[c]{@{}l@{}}$\bullet$ fewer types of risks, free from the\\ \ \ \ single point failure problem~\cite{fan2021dr}\end{tabular}                                                                 & \begin{tabular}[c]{@{}l@{}}$\bullet$ specific threats, e.g., breaking network access,\\ \ \ \ unknown stakeholders or determining resource\\ \ \ \  locations, single point failure problem~\cite{mahadevappa2021review}\end{tabular}                                  \\ \hline
\multirow{7}{*}{\textbf{\rotatebox{90}{Approach}}} & \multirow{7}{*}{\textbf{\rotatebox{90}{Related}}} & Assumption & \begin{tabular}[c]{@{}l@{}}$\bullet$ Almost enough resources support \\ \ \ \ complex inspection schemes.\end{tabular}    & \begin{tabular}[c]{@{}l@{}}$\bullet$ Limited resources can not conduct complex\\ \ \ \ inspection tasks.\end{tabular}                                                                                                                                            \\ \cline{3-5}  
                                              &                                       & Trust model                                & $\bullet$ CIPs ensure data integrity.                                                                               & $\bullet$ EIPs are responsible for data integrity.                                                                                                                                               \\ \cline{3-5}  
                                              & 
                                                                                     & Main function                              & $\bullet$ verification                                                                                                         & $\bullet$ verification, localization, and geo-assurance                                                                                                                                     \\ \cline{3-5}  
                                              & 
                                                                                     & Main feature                               & $\bullet$ less limitation                                                                                                      & $\bullet$ more lightweight for every participant                                                                                                                                                \\ \cline{3-5}  
                                              & 
                                                                                     & Inspection way                             & $\bullet$ one-by-one inspection                                                                                                 & $\bullet$ inspection in a parallel way                                                             \\ \cline{3-5}  
                                              &                                               & Inspection frequency                       & $\bullet$ less, having centralized control                                                                    & \begin{tabular}[c]{@{}l@{}}$\bullet$ more due to a higher level of threats,\\ \ \ \ lack of centralized control  \end{tabular}

\\ \hline \hline
\end{tabular}}
\end{table}

\subsection{Edge Data Integrity Verification Versus Cloud Data Integrity Verification}
\label{subsec:Edge Data Integrity Versus Cloud Data Integrity}
The detailed comparison between EDIV and CDIV is summarized in Table~\ref{tab:Cloud Data Integrity vs Edge Data Integrity}. In brief, there are three key distinctions. First, \emph{edge nodes as data carriers lead to device-related discrepancies including the number, location, capacity, and providers of devices.} In this respect, edge nodes' protection systems are more brittle than those of cloud servers, and therefore a variety of attacks that could be ineffectual against cloud servers can seriously endanger the integrity of data cached on edge~\cite{ranaweera2021survey, corcoran2016mobile}. Second, \emph{data cached on edge suffers from a greater diversity of attacks.} Obviously, attack diversification hugely raises the corruption undetected probability~\cite{mukherjee2020intelligent}. Because of it, even if corruption occurs, most data owners/users might not be able to notice it. Third, \emph{diverse question scenarios bring in different problem-solving approaches.} The challenges of creating an integrity verification mechanism for edge computing are directly attributed to this aspect. The majority of CDIV schemes in use today are coarse-grained~\cite{liu2015external}, which is unfit for edge computing because of the more complex systems and applications. Fine-grained and lightweight integrity verification approaches are necessitated in edge computing environments~\cite{alzoubi2021fog}.

\begin{figure}
    \centering
	\begin{minipage}[h]{0.35\linewidth}
		\includegraphics[width=1\linewidth]{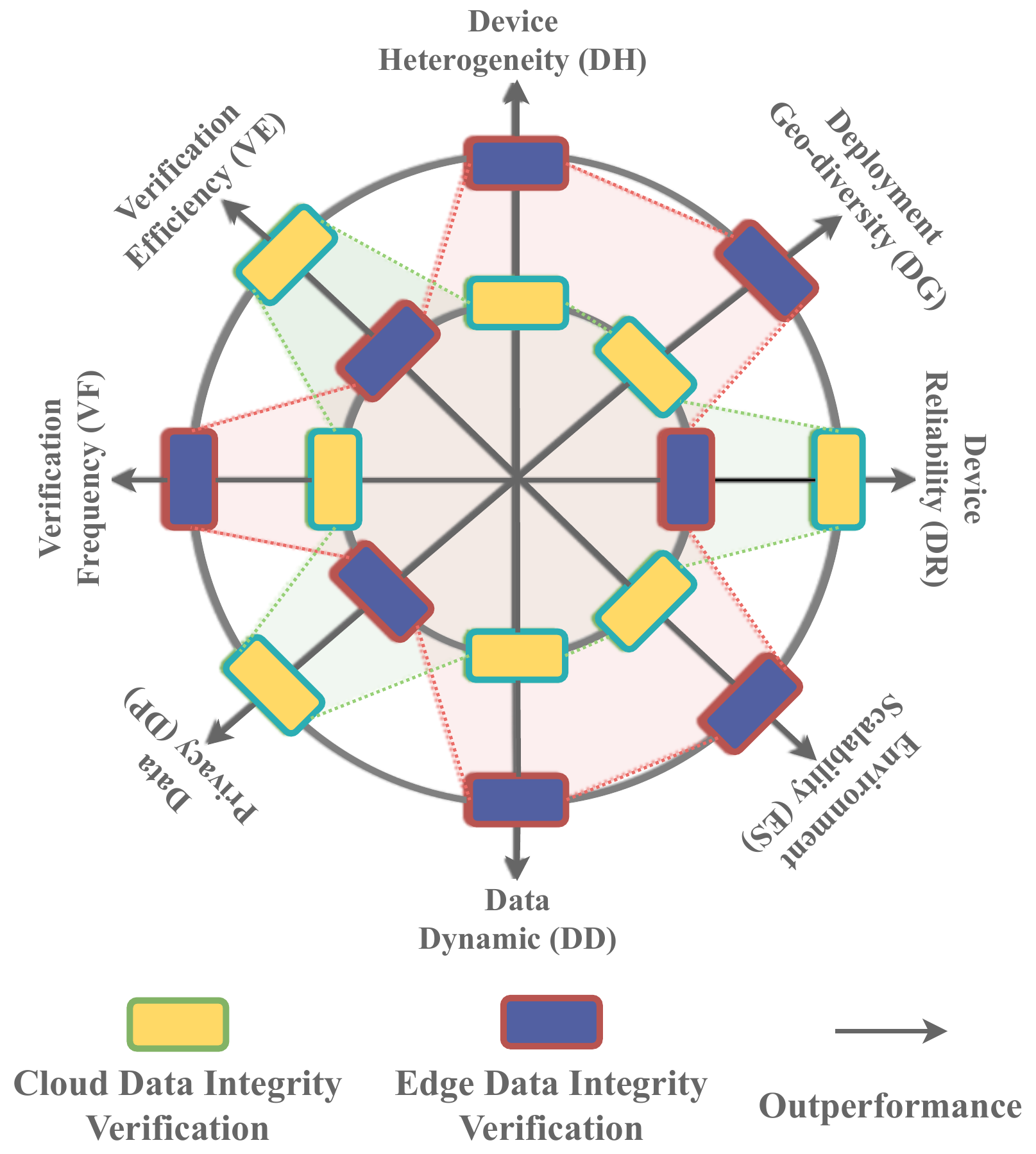}
	\end{minipage}\hspace{-1 mm}
			\begin{minipage}{0.65\textwidth}
			\centering
\small
\renewcommand\arraystretch{1.2}
\begin{threeparttable}
\begin{tabular}{c|ll}
\hline\hline
Features & \multicolumn{1}{c}{CDIV$^1$}                             & \multicolumn{1}{c}{EDIV$^1$}                     \\ \hline
DH~\cite{zhang2017computing}       & $\Box$ similar data centers                            & $\blacksquare$ inequable edge nodes \\ \hline
DG~\cite{zhang2017computing}       & $\Box$ one or several locations                     & $\blacksquare$ numerous locations           \\ \hline
DR~\cite{liu2020reliability}       & $\blacksquare$ strong security guarantee   & $\Box$ weak assurance mechanism      \\ \hline
ES~\cite{ren2019survey}       & $\Box$ change infrequent                          & $\blacksquare$ easy extension               \\ \hline
DD~\cite{zhang2020edge}       & $\Box$ steady data                           & $\blacksquare$ dynamic data         \\ \hline
DP~\cite{xiao2019edge}       & $\blacksquare$ powerful privacy protection                    & $\Box$ weak protection policy  \\ \hline
VF~\cite{li2020auditing}       & $\Box$ low-frequency                      & $\blacksquare$ high-frequency         \\ \hline
VE~\cite{li2020auditing}       & $\blacksquare$ high-efficiency  & $\Box$ low-efficiency   \\ \hline \hline
\end{tabular}
\begin{tablenotes}
		\item 1: $\blacksquare$ high-performance; $\Box$ low-performance.
     \end{tablenotes}
\end{threeparttable}
		\end{minipage}
    \caption{A comparative visual summary between edge data integrity verification and cloud data integrity verification}
	\label{fig:A comparative visual summary between cloud data integrity and edge data integrity}
 \vspace{-1 em}
 \end{figure}

\par We further state the key differences from a high-level point of view in Fig.~\ref{fig:A comparative visual summary between cloud data integrity and edge data integrity}, where each of the edges indicates one of the features that cloud and edge exhibit discrepancy in data integrity. It can be deduced from the radar graph and the table that edge nodes are more heterogeneous and follow geo-diversity deployments~\cite{zhang2017computing}. Plus, data privacy leakage during integrity verification occurs more often in edge because it broadens the real-world attack surface from the perspective of weak computation power, attack unawareness, protocol heterogeneity, and coarse-grained access control~\cite{xiao2019edge}. Furthermore, data dynamic issues should be considered more rigorously in edge, as data cached on edge is changed faster than on cloud~\cite{zhang2020edge}. Besides, since edge nodes have less reliability~\cite{liu2020reliability}, verification frequency in edge domains should be higher. Finally, verification efficiency currently achievable is hard to match existing CDIV solutions due to highly scalable edge environments.

\subsection{System Models and Key Processes}
\label{subsec:System Models and Key Processes}
\begin{figure*}[!tp]
\centering
 \setlength{\abovecaptionskip}{0pt}%
\setlength{\belowcaptionskip}{0pt}%
    \subfigure[Private audit process]{
    \label{fig:Data Owner-edge nodes}
        \begin{minipage}[t]{0.32\textwidth}
        \includegraphics[width=1\textwidth]{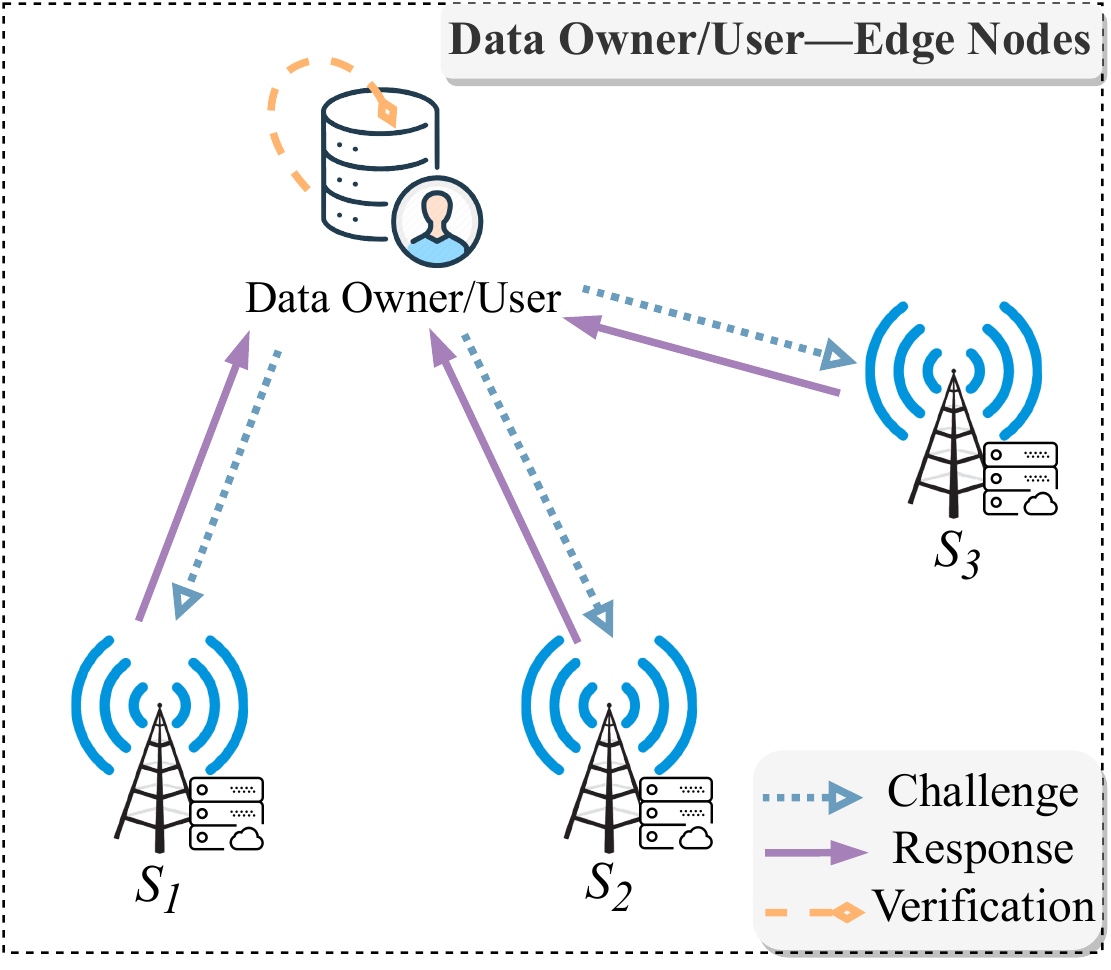}
        \end{minipage}}
    \subfigure[Public audit process]{
    \label{fig:TPA-edge nodes}
        \begin{minipage}[t]{0.32\linewidth}
        \includegraphics[width=1\textwidth]{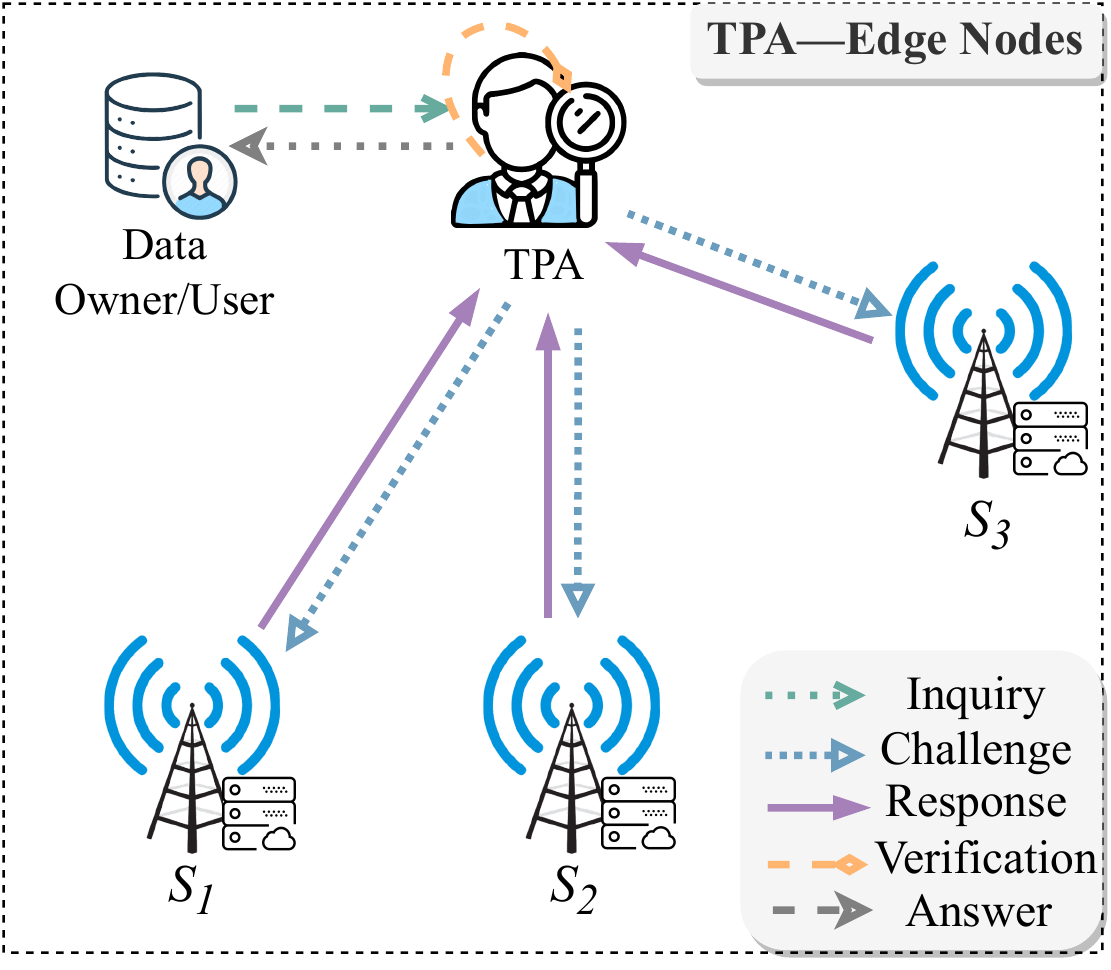}
        \end{minipage}}
    \subfigure[Cooperative audit process]{
    \label{fig:Edge Server-Edge Server}
        \begin{minipage}[t]{0.32\linewidth}
        \includegraphics[width=1\textwidth]{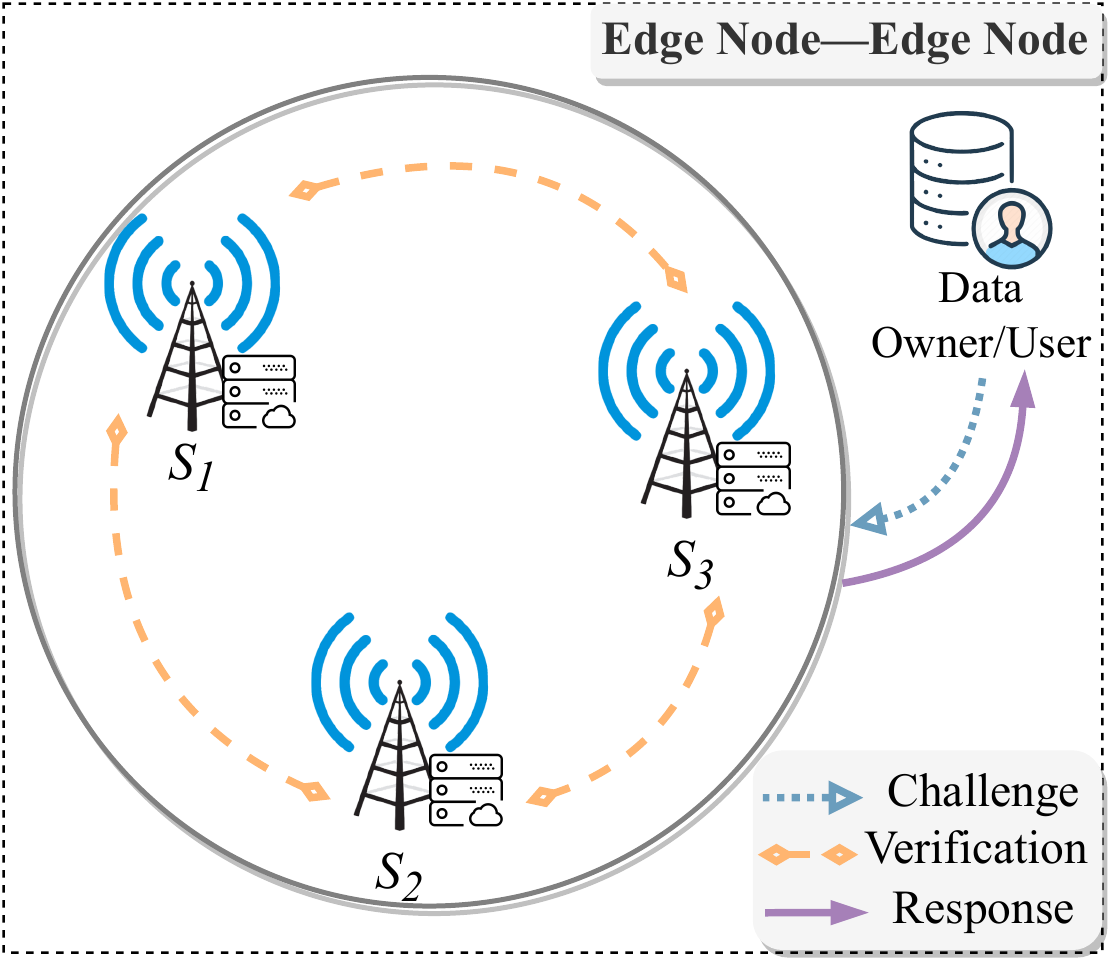}
        \end{minipage}}
\caption{System models. There are three types of system models: (1) private audit where data owner/user interacts with edge nodes; (2) public audit where TPA interacts with edge nodes; and (3) cooperative audit involving interactions among edge nodes.}
\label{fig:system models}
\vspace{-1em}
\end{figure*}

EDIV problems usually happen in data-driven services that take advantage of edge computing architectures~\cite{zhang2017cooperative, yao2019mobile, su2018edge}, where edge nodes are always expected to be involved in the process of integrity verification for the ease of security concerns of both data owners and data users~\cite{chang2014bringing}. Due to the limitations of processing and storage capacities of data owners/users, sometimes the integrity check is performed by a \textbf{third party auditor (TPA)}~\cite{tong2019privacy}. To sum up, there are four types of entities related to the EDIV problem:
\begin{itemize}
  \item
    \textbf{Data Owner (DO)}: It could be an Application vendor, developer, etc. It outsources its latency-sensitive data on multiple geographically distributed edge nodes to serve nearby DUs, as shown in Fig.~\ref{fig:System model}. To enhance user experience, it queries data and requests for data integrity verification on edges.
  \item
    \textbf{Data User (DU)}: It may be an IoT device, mobile subscriber, etc. It would check the integrity of queried data, because of self-interest concerns, by the interaction with edge nodes like the case presented in Fig.~\ref{fig:Data Owner-edge nodes} and Fig.~\ref{fig:TPA-edge nodes}.
  \item
    \textbf{Edge Node (EN)}: It could be an access point, base station, etc. As illustrated in Fig.~\ref{fig:System model}, it caches data replicas for DOs, can be reached by nearby DUs with low access latency, and assists in the completion of integrity checks.
  \item
    \textbf{Third Party Auditor (TPA)}: It could be an agency, certification body, etc. It performs an external and independent audit of the integrity of data cached on remote ENs for DOs/DUs, as displayed in Fig.~\ref{fig:TPA-edge nodes}. Typically, TPAs have powerful computing and storage capabilities.
\end{itemize}

\par Generally, a system model defines the purpose and context for approach usage. More specifically, it describes what kind of participant is involved and what assumption is held~\cite{ahmed2018survey}. Notably, not all of the entities mentioned above are included in every existing EDIV solution. We extract three generic system models in brief, including private audit, public audit, and cooperative audit. These models mainly differentiate on what participants are engaged. In private audit, it is the data owner/user that verifies EDI. However, the data owner/user is not totally trustworthy, as it may deliberately claim incorrect verification results to obtain monetary or service compensation from EIPs~\cite{huang2021profit}. Thus, \emph{from the edge nodes' perspective, it is impossible to determine whether verification results are reliable or not}. To eliminate this concern and alleviate the verification burden on data owners/users, public audit occurs, in which TPA acting as a trusted agency fulfills integrity inspection~\cite{xu2022secure, ge2021revocable}. On the downside, \emph{the security of EDIV is hard to guarantee if TPA is byzantine.} Very recently, cooperative audit comes into follow-up researcher's sight to improve verification fairness. In this case, blockchain or other distributed technologies are adopted to enable edge nodes to collectively finish inspection tasks without the participation of TPA or even data owners/users~\cite{zhao2020blockchain, chen2018stochastic, wang2021rdic}. We introduce these three system models with key processes as follows.

\subsubsection{Private Audit}
As shown in Fig.~\ref{fig:Data Owner-edge nodes}, a data owner/user inspects outsourced data using a \emph{Challenge-Response} mechanism, in which it interacts with multiple edge nodes via message exchange, and EDI can be ensured with a high probability if edge nodes provide correct integrity proofs~\cite{li2020auditing, li2021inspecting}. The key process of private audit is as follows.
\begin{itemize}
  \item
    \textbf{Challenge(.) \bm{$\rightarrow$} \bm{$chal.message(.)$}}: DO/DU runs this randomized algorithm to build and send a challenge message $chal.message(.)$ to each EN.
  \item
    \textbf{Response\bm{$(d,\ chal.message(.))$}\bm{$\rightarrow$} \bm{$resp.message(r)$}}: This randomized algorithm is run by each EN to prove the integrity of cached data replicas $d$. It takes as input the challenge message $chal.message(.)$ and cached data replicas $d$, and then returns a response message $resp.message(r)$ with integrity proof $r$.
   \item
    \textbf{Verification\bm{$(\cup\{resp.message(r)\})$}\bm{$\rightarrow$} \bm{$(\top, \bot[, \{id\}])$}}: DO/DU runs this deterministic algorithm. It takes as input the set of received response messages and determines whether the corresponding data replicas are integral. If proofs are verified in a batch way, it further locates corrupted data replicas.
\end{itemize}

\subsubsection{Public Audit}
As presented in Fig.~\ref{fig:TPA-edge nodes}, a data owner/user authorizes TPA to complete EDIV for ease of the verification burden~\cite{tong2022privacy, tong2019privacy}. In this case, the data owner/user sends an inquiry message to TPA, and then TPA is responsible for checking data integrity, following the same \emph{Challenge-Response} mechanism. After finishing it, TPA returns verification results back to the data owner/user. The key process of public audit is as follows.
\begin{itemize}
  \item
    \textbf{Inquiry(.)\bm{$\rightarrow$} \bm{$inq.message(.)$}}: This algorithm is run by DO/DU. It outputs an inquiry message $inq.message(.)$.
  \item
    \textbf{Challenge}\bm{$(inq.message(.))$}\bm{$\rightarrow$} \bm{$chal.message(.)$}: TPA runs this randomized algorithm to build and send a challenge message $chal.message(.)$ to each EN.
  \item
    \textbf{Response\bm{$(d,\ chal.message(.))$}\bm{$\rightarrow$} \bm{$resp.message(r)$}}: This randomized algorithm is run by each EN to prove the integrity of cached data replicas $d$. It inputs the challenge message $chal.message(.)$ and cached data replicas $d$, and sends back a response message with an integrity proof $r$.
   \item
    \textbf{Verification\bm{$(\cup\{resp.message(r)\})$}\bm{$\rightarrow$} \bm{$(\top, \bot[, \{id\}])$}}: TPA runs this deterministic algorithm. It takes as input the set of response message $\cup\{resp.message(r)\}$, and determines whether the corresponding data replicas are integral.
   \item
    \textbf{Answer\bm{$(\top, \bot[, \{id\}])$}\bm{$\rightarrow$} \bm{$ans.message(.)$}}: The algorithm is run by TPA. It takes as input the verification result, and sends an answer message $ans.message(.)$ to DO/DU.
\end{itemize}

\begin{table}[!tp]
\small
\renewcommand\arraystretch{1.2}
\caption{Review on existing internal attacks regarding EDIV}
\label{tab:Review on existing internal attacks regarding the EDIV verification}
\setlength{\tabcolsep}{2mm}{
\begin{tabular}{c|l|c}
\hline\hline
\textbf{Attack Type} & \textbf{Description}  & \textbf{Potential Context}                                                                    \\ \hline
Spoofing Attack~\cite{xu2016probabilistic}       & \begin{tabular}[c]{@{}l@{}}$\bullet$ Dishonest DO/DU/TPA checks the received proofs and claims\\ \ \ \  an incorrect verification result. \end{tabular}       &  Public and private audits               \\ \hline
Replay Attack~\cite{yi2017efficient}       & \begin{tabular}[c]{@{}l@{}}$\bullet$ Dishonest EN deduces the latest integrity proof from previously\\ \ \ \  generated ones. \end{tabular}    &  All system models     \\ \hline
Forgery Attack~\cite{ping2020public}      & \begin{tabular}[c]{@{}l@{}}$\bullet$ Dishonest EN forges an integrity proof to bypass the integrity\\ \ \ \  check. \end{tabular}     &  All system models                     \\ \hline
Replace Attack~\cite{li2022identity}       & \begin{tabular}[c]{@{}l@{}}$\bullet$ Dishonest EN replaces a damaged block with another intact\\ \ \ \  block saved by itself to try to pass the verification.  \end{tabular}       &  All system models               \\ \hline
Data Leakage Attack~\cite{ren2021integrity}       & \begin{tabular}[c]{@{}l@{}}$\bullet$ Dishonest TPA deduces outsourced data during the verification\\ \ \ \  protocol.  \end{tabular} &  Public audit \\ \hline
Outsourcing Attack~\cite{benet2017proof}   & \begin{tabular}[c]{@{}l@{}}$\bullet$ Dishonest EN intercepts integrity proofs created by other ENs\\ \ \ \  as its own proof.  \end{tabular} &  Cooperative audit            \\ \hline
Byzantine Attack~\cite{meena2013survey}     & \begin{tabular}[c]{@{}l@{}}$\bullet$ Dishonest EN tampers with honest ENs' integrity proofs when\\ \ \ \  returning the integrity proof. \end{tabular} &  Cooperative audit    \\ \hline
Collusion Attack~\cite{gaetani2017blockchain}     & \begin{tabular}[c]{@{}l@{}}$\bullet$ Dishonest ENs collude together to corrupt the cached data to\\ \ \ \  deceive DO/DU/TPA. \end{tabular} &  Cooperative audit \\ \hline\hline
\end{tabular}}
\end{table}

\subsubsection{Cooperative Audit}
In the last two years, the mentality of EDIV problem-solving has been further enlarged. Edge nodes check data integrity by themselves without interaction with a data owner/user or TPA~\cite{li2021cooperative,li2022edgewatch}, as illustrated in Fig.~\ref{fig:Edge Server-Edge Server}. In this case, edge nodes usually adopt consensus algorithms to reach an agreement about the integrity status of cached data replicas, and directly return the verification result that all agree on to the data owner/user. The key process is demonstrated as follows.
\begin{itemize}
   \item
    \textbf{Challenge\bm{$(.)$}\bm{$\rightarrow$} \bm{$chal.message(.)$}}: DO/DU runs this deterministic algorithm to build and send a challenge message $chal.message(.)$ to ENs.
   \item
    \textbf{Verification\bm{$(chal.message(.))$}\bm{$\rightarrow$} \bm{$(\top, \bot[, \{id\}])$}}: ENs collaboratively run this deterministic algorithm. It inspects cached data replicas and locates corrupted ones.
  \item
    \textbf{Response\bm{$(\top, \bot[, \{id\}])$}\bm{$\rightarrow$} \bm{$resp.message(.)$}}: This algorithm is run by one or multiple ENs to return the verification result to DO/DU.
\end{itemize}

\par We further identify a number of EDIV-related attacks that are targeted to damage verification processes and launched by EDIV participants (i.e., DO, DU, EN, and TPA) in various system models, as summarized in Table~\ref{tab:Review on existing internal attacks regarding the EDIV verification}. Please note that there is a strong correlation between those attacks and security assumptions adopted in solutions. For example, if TPA is supposed to be fully trustworthy, data leakage attacks never happen in the verification process.

\section{Evaluation Criteria on Verification Approach}
\label{sec:Evaluation Criteria on Verification Approaches}
In this section, we explore several evaluation criteria to uncover the strengths and weaknesses of existing EDIV solutions. This set of criteria is also utilized to assess the effectiveness of verification approaches. A brief classification of the criteria that would be covered in this survey and the methodologies are illustrated in Fig.~\ref{fig:Evaluation Criteria} to highlight the relationship among these indicators, followed by detailed descriptions below.

\par First, substantial effort is necessary to ensure well-defined, comprehensive, and coherent problem scenarios. Existent EDIV solutions can be categorized into two tiers from the scenario support perspective: multi-replica and multi-owner. Notably, in edge domains, we usually do not investigate single-replica storage cases (i.e., DO caches one data replica on a single EN) because of two reasons: (1) existing CDIV solutions could be directly applied to handle the EDIV problem with imperceptible modification; (2) single-replica storage is unable to support low-latency services for geographically distributed DUs, which deviates from the objective of EC~\cite{mor2019global}. 

\par Specifically, multi-replica support represents that the solution is designed for and works well in multi-replica scenarios, where a single DO deploys its data replicas across multiple geo-distributed ENs to serve various users in different regions, like the example presented in Fig.~\ref{fig:System model}. Supporting multi-replica is one of the fundamental characteristics of effective EDIV solutions. Moreover, multi-owner support denotes that the solution can simultaneously inspect multiple data replicas for multiple data owners. While it may seem intuitive to extend solutions designed for multi-replica scenarios to accommodate multi-owner cases, this is unfeasible in reality. If multi-replica solutions are trivially extended to support multi-owner with data integrity assurance, each data owner has to perform the same verification workflow to interact with all corresponding edge nodes. In particular, the edge node that caches numerous data replicas from various data owners may suffer high computation and communication overheads, as it has to process multiple audit requests from dissimilar data owners at the same time~\cite{ji2020inspection}. Clearly, these trivial extensions could incur a new performance bottleneck and a tremendous workload on edge nodes. Thus, digging into multi-owner scenarios to design appropriate EDIV approaches is meaningful and critical in practice.

\par On top of the specific scenarios, \emph{security assumption} serves as the foundation when designing EDIV solutions and is intricately related to the classified criteria~\cite{li2017fuzzy}. Generally, security assumptions are linked to four entities: data owners, data users, edge nodes, and TPAs. Each of them can be assumed to be \emph{untrusted}, \emph{semi-trusted}, or \emph{trusted}. Different assumptions lead to various cases and need different approaches to handle them. For example, if TPA is supposed to be fully trusted, there is no need to develop a privacy-preserving EDIV method. In contrast, if this trust assumption does not hold, privacy issues should be jointly considered. Next, we elaborate on specific indicators.

\subsection{Efficiency Related Indicators}
\label{subsec:Efficiency Related Indicators}
EDIV efficiency consists of three aspects including computation, storage, and communication, which further derives three indicators, i.e., batch support for computation efficiency improvement, blockless verification for communication efficiency improvement, and stateless verification for storage efficiency improvement. We articulate them separately as follows.

\subsubsection{Batch Support (BS)}
\label{subsubsec:Batch Support}
Batch support refers to that a method could inspect the integrity of multiple data replicas simultaneously~\cite{kai2013efficient}. It is a general indicator for both multi-replica and multi-owner scenarios. Its main purpose is to improve computation efficiency so that data replicas can be inspected more frequently over a fixed time interval. To date, lots of related work, e.g.,~\cite{tong2019privacy, tong2022privacy}, supports batch verification, especially in edge computing environments.

\subsubsection{Blockless Verification (BV)}
\label{subsubsec:Blockless Verification}
Blockless verification refers to that an approach should not ask data owners/users or TPAs to retrieve outsourced data replicas from the remote edge nodes for verification purposes~\cite{lin2017data}. This is a prerequisite for all EDIV solutions, since it is an unnecessary and communication-consuming task to access the whole data replicas (particularly with big sizes) cached on edge nodes and check integrity. In blockless verification, it is small-sized integrity proofs (e.g., hash strings) that are generated and transferred to the data owner/user or TPA for proof verification, which reduces communication overhead fundamentally.

\subsubsection{Stateless Verification (SV)}
\label{subsubsec:Stateless Verification}
Stateless verification means that neither a data owner/user/TPA nor an edge node is required to cache previous verification results in order to perform future audits~\cite{ferretti2018symmetric}. In short, every challenge request from data owners/users or TPAs is time-independent, which aims to save storage space for each side. This is an indirect requirement of data integrity methods. Otherwise, it may result in a situation in which keeping prior audit states becomes a storage burden for each participant.

\begin{figure}[!tp]
    \centering
    \includegraphics[width=0.9\linewidth]{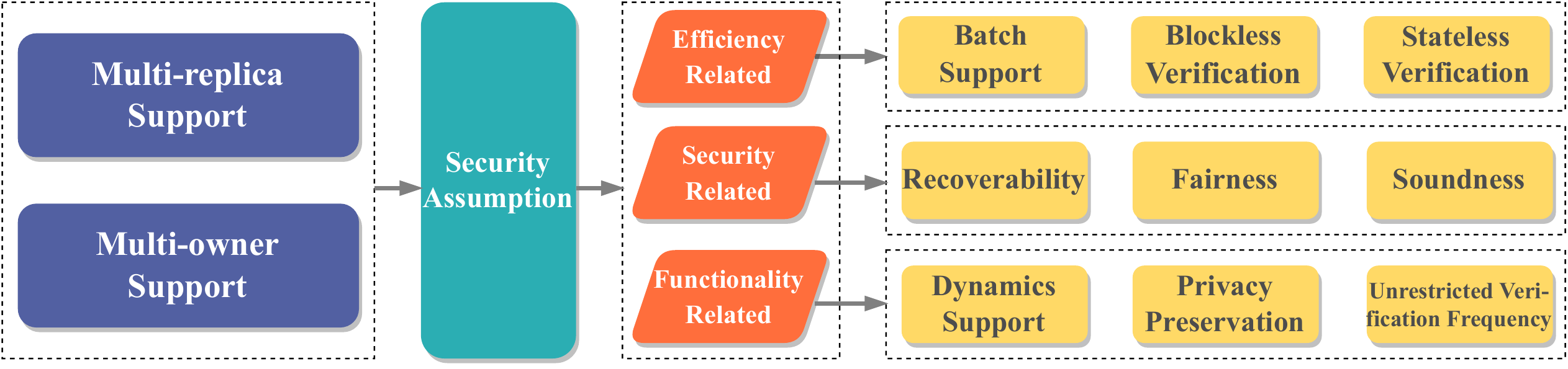}
    \caption{Evaluation criteria of edge data integrity verification solutions}
    \label{fig:Evaluation Criteria}
    \vspace{-1em}
\end{figure}
 
\subsection{Security Related Indicators}
\label{subsec:Security Related Indicators}
In addition to efficiency-related metrics, a number of security attributes should be taken into consideration when designing an appropriate EDIV approach. From a high-level point of view, an EDIV scheme would be more practical if it could recover the corrupted data replica after identifying it. Furthermore, fairness is another important indicator, as keeping fair is the basis of incentives that motivate edge nodes to act normal and perfect. Last, soundness ensures that EDIV solutions can secure against various types of attacks during verification, yielding a correct verification result. Next, we introduce them in detail.

\subsubsection{Recoverability (Re)}
\label{subsubsec:Recoverability}
Recoverability refers to that an EDIV solution can not only find out the corrupted data replicas but also complete data recovery to avoid an unpleasant situation~\cite{chen2011rcda}. This property fills the gaps left by corruption so that protects EIPs' reputation and guards data owners to keep user experience.

\subsubsection{Fairness (Fa)}
\label{subsubsec:Fairness}
Fairness is displayed in two aspects. First, from the edge nodes' perspective, a satisfactory solution should ensure that data owners/users can not deliberately assert incorrect verification results~\cite{saxena2016cloud}, i.e., edge nodes are capable of telling whether or not inspection results can be trusted. Accordingly, malicious data owners/users are impossible to damage the reputation of EIPs. Second, from data owners'/users' perspective, a feasible solution should provide protection against legitimate but malicious edge nodes who may collude to obtain a misleading verification result~\cite{zikratov2017ensuring}, especially in cooperative audit. Unequivocally, fairness is a key indicator for both edge nodes and data owners/users~\cite{shen2018enabling}.

\subsubsection{Soundness (So)}
\label{subsubsec:Soundness}
Soundness refers to that an edge node is unable to pass verification unless it provides a correct integrity proof~\cite{zhu2012efficient}. If an edge node can pass a challenge request without holding the data or with corrupted data, the data owner/user is incapable of detecting data corruption in a timely manner, resulting in potentially far-reaching business ramifications. Therefore, the soundness property guarantees data reliability and is a necessity for approach design.

\subsection{Functionality Related Indicators}
\label{subsec:Functionality Related Indicators}
EDIV is a wide-range problem, and solutions may have lots of appendant sub-functions besides integrity verification and corruption localization, such as dynamic verification, privacy preservation, and unrestricted verification frequency. Obviously, the more the EDIV approach supports, the better it is.

\subsubsection{Dynamic Verification (DV)}
\label{subsubsec:Dynamic Verification}
Dynamic verification refers to that an EDIV approach can work steadily when the cached data replicas are updated by data owners~\cite{zhang2016efficient}. It is an important indicator, as data dynamics is a fundamental characteristic and often occurs in edge computing environments. An approach supporting dynamic verification property would be more practical, especially in industry.

\subsubsection{Privacy Preservation (PP)}
\label{subsubsec:Privacy Preservation}
Privacy preservation requires that TPA has no personal knowledge of the sensitive information of data owners/users, data replicas, and edge nodes while yet validating the integrity of outsourced data~\cite{hao2011privacy}. Privacy leakage often occurs in public audit owing to the curious TPA involvement, but it is uncommon in private or collaborative audits.

\subsubsection{Unrestricted Verification Frequency (UVF)}
\label{subsubsec:Unrestricted Frequency}
Unrestricted verification frequency implies that there should be no limits on the number of challenges issued by a data owner/user or TPA for integrity validation~\cite{khedr2019cryptographic}. It is also known as unbounded inquiries. EDIV is a continuous process, in which a data owner/user or TPA runs the verification procedure at regular intervals to detect data corruption. The computation efficiency of the data integrity completion has a direct impact on the frequency of challenge requests. If the verification procedure of a data integrity method is computationally demanding, the data owner/user or TPA would adopt it less frequently, and consequently, unrestricted verification frequency will suffer.

\section{Existing Edge Data Integrity Verification Solutions}
\label{sec:Existing EDIV solutions}
In this section, we summarize the development process of EDIV problems and then survey the literature advances. We explore the following databases: Web of Science, Google Scholar, IEEE Xplore, and ACM library to search papers based on the keywords: edge data integrity, integrity attack, edge computing, data security, and integrity in edge. Adopting the criteria presented in Fig.~\ref{fig:Evaluation Criteria}, we qualitatively review existing EDIV approaches and delineate their respective strengths and weaknesses. For ease of understanding and interpretation, we list EDIV approaches in the taxonomy Table~\ref{tab:Summary and Qualitative Comparison of Existing Solutions} to summarize the qualitative aspects. Further, we articulate the key contributions, limitations, and evaluation metrics of each reference work in Table~\ref{tab:Pros and Cons of Private Audit}, Table~\ref{tab:Pros and Cons of Public Audit}, and Table~\ref{tab:Pros and Cons of Cooperative Audit} for private audit, public audit, and cooperative audit, respectively.

\par
\begin{figure}[!tp]
    \centering
    \includegraphics[width=1\linewidth]{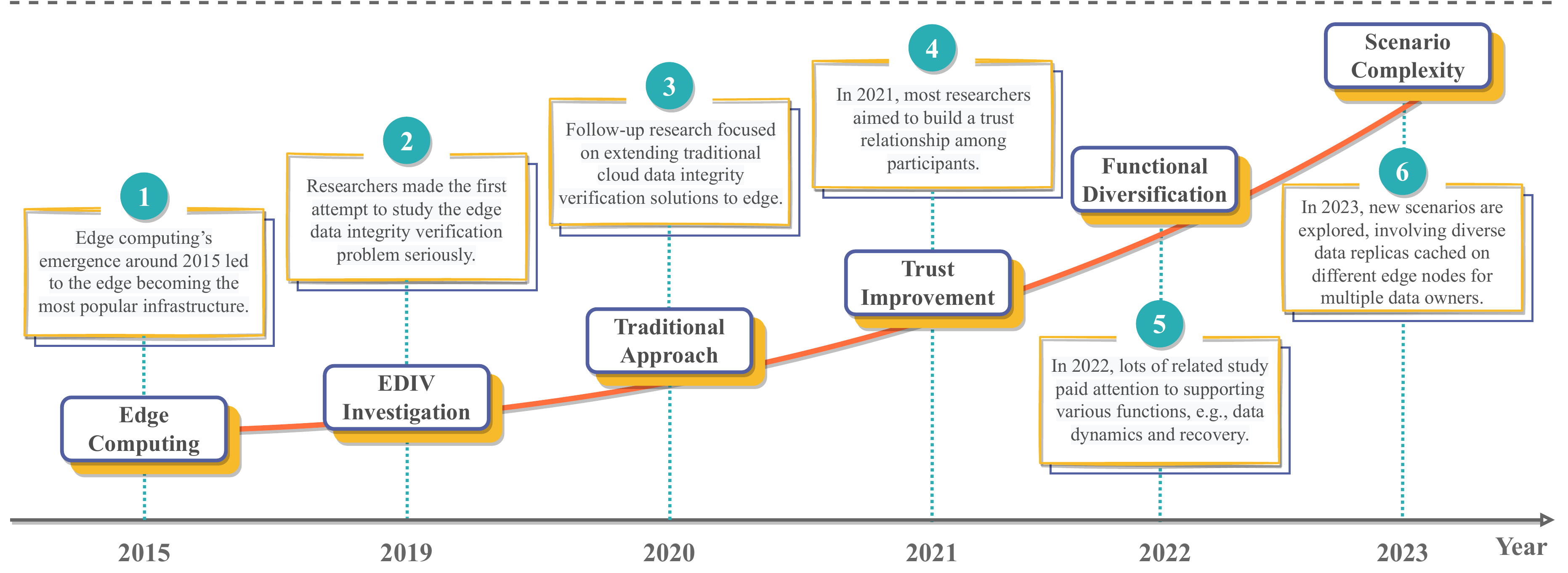}
    \caption{Development timeline of edge data integrity investigation}
    \label{fig:timeline}
    \vspace{-1 em}
\end{figure}

 Overall, EDIV is a novel problem and is still in its early stages. This problem has evolved over five years since its inception. Fig.~\ref{fig:timeline} depicts a schematic layout of the EDIV problem development process. It has made attractive progress and achievements over the past few years but also endured frustrations and setbacks. In the following, we illustrate several stages of the evolution of EDIV solutions.

\begin{table*}[]
\small
\renewcommand{\arraystretch}{1.2}
\caption{Qualitative Comparison of Existing Solutions}
\label{tab:Summary and Qualitative Comparison of Existing Solutions}
\setlength{\tabcolsep}{2mm}{
\begin{threeparttable}
\begin{tabular}{m{1.5cm}<{\centering}|m{0.6cm}<{\centering}|m{1cm}<{\centering}|m{1.8cm}<{\centering}|m{1cm}<{\centering}m{1cm}<{\centering}m{0.5cm}<{\centering}|m{1cm}<{\centering}m{1cm}<{\centering}m{0.5cm}<{\centering}|m{1.2cm}<{\centering}m{1.2cm}<{\centering}m{0.5cm}<{\centering}}
\hline \hline
\multirow{2}{*}{\textbf{Category}}                                             & \multirow{2}{*}{\textbf{Ref.}} &  \multicolumn{1}{c|}{\multirow{2}{*}{\textbf{Scenario$^1$}}} & \multirow{2}{*}{\textbf{\begin{tabular}[c]{@{}c@{}}Security\\ Assumption$^2$\end{tabular}}} & \multicolumn{3}{c|}{\textbf{Effi. Related$^3$}}                                                       & \multicolumn{3}{c|}{\textbf{Secur. Related$^3$}}                                                         & \multicolumn{3}{c}{\textbf{Funct. Related$^3$}}                                                     \\ \cline{5-13} 
                                                                               &                                &                                                                                            & \multicolumn{1}{c|}{}                                   & \multicolumn{1}{c|}{\textbf{BS}} & \multicolumn{1}{c|}{\textbf{BV}} & \multicolumn{1}{c|}{\textbf{SV}} & \multicolumn{1}{c|}{\textbf{Re}} & \multicolumn{1}{c|}{\textbf{Fa}} & \multicolumn{1}{c|}{\textbf{So}} & \multicolumn{1}{c|}{\textbf{DV}} & \multicolumn{1}{c|}{\textbf{PP}} & \multicolumn{1}{c}{\textbf{UVF}} \\ \hline
\multirow{10}{*}{\begin{tabular}[c]{@{}c@{}}Private\\ Audit\end{tabular}}       & \cite{li2020auditing}          & \multicolumn{1}{c|}{O-E}                                                                 &   TO-UE                                    & \multicolumn{1}{c|}{\begin{minipage}[b]{0.02\columnwidth}
		\centering\raisebox{-.2\height}{\includegraphics[width=\linewidth]{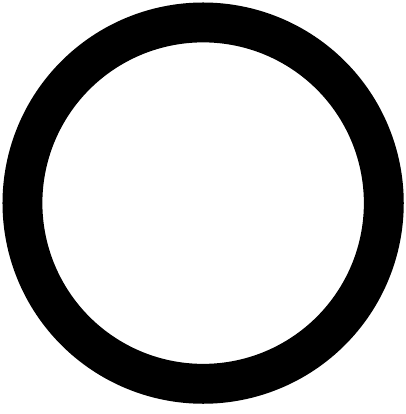}}
	\end{minipage}}            & \multicolumn{1}{c|}{\begin{minipage}[b]{0.02\columnwidth}\centering\raisebox{-.2\height}{\includegraphics[width=\linewidth]{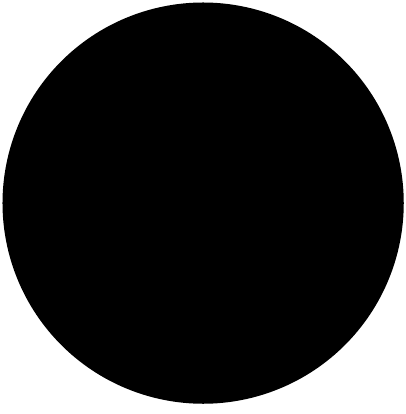}}
	\end{minipage}}            &    \begin{minipage}[b]{0.02\columnwidth}\centering\raisebox{-.2\height}{\includegraphics[width=\linewidth]{fig/icon/circle.pdf}}
	\end{minipage}                              & \multicolumn{1}{c|}{\begin{minipage}[b]{0.02\columnwidth}\centering\raisebox{-.2\height}{\includegraphics[width=\linewidth]{fig/icon/annulus.pdf}}
	\end{minipage}}            & \multicolumn{1}{c|}{\begin{minipage}[b]{0.02\columnwidth}\centering\raisebox{-.2\height}{\includegraphics[width=\linewidth]{fig/icon/annulus.pdf}}
	\end{minipage}}            &   \begin{minipage}[b]{0.02\columnwidth}\centering\raisebox{-.2\height}{\includegraphics[width=\linewidth]{fig/icon/circle.pdf}}
	\end{minipage}                               & \multicolumn{1}{c|}{\begin{minipage}[b]{0.02\columnwidth}\centering\raisebox{-.2\height}{\includegraphics[width=\linewidth]{fig/icon/annulus.pdf}}
	\end{minipage}}            & \multicolumn{1}{c|}{\begin{minipage}[b]{0.02\columnwidth}\centering\raisebox{-.2\height}{\includegraphics[width=\linewidth]{fig/icon/circle.pdf}}
	\end{minipage}}            &   \begin{minipage}[b]{0.02\columnwidth}\centering\raisebox{-.2\height}{\includegraphics[width=\linewidth]{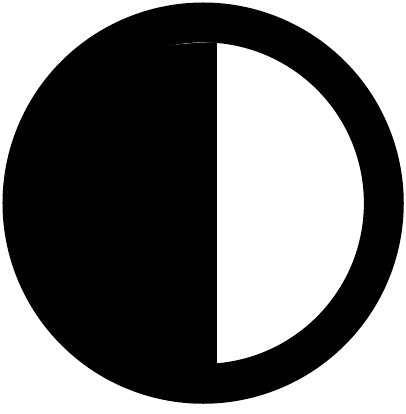}}
	\end{minipage} \\ \cline{2-13} 
    & \cite{li2021inspecting}        &\multicolumn{1}{c|}{O-E}  &  TO-UE                   & \multicolumn{1}{c|}{\begin{minipage}[b]{0.02\columnwidth}\centering\raisebox{-.2\height}{\includegraphics[width=\linewidth]{fig/icon/circle.pdf}}
	\end{minipage}}            & \multicolumn{1}{c|}{\begin{minipage}[b]{0.02\columnwidth}\centering\raisebox{-.2\height}{\includegraphics[width=\linewidth]{fig/icon/circle.pdf}}
	\end{minipage}}            &  \begin{minipage}[b]{0.02\columnwidth}\centering\raisebox{-.2\height}{\includegraphics[width=\linewidth]{fig/icon/circle.pdf}}
	\end{minipage}                                & \multicolumn{1}{c|}{\begin{minipage}[b]{0.02\columnwidth}\centering\raisebox{-.2\height}{\includegraphics[width=\linewidth]{fig/icon/annulus.pdf}}
	\end{minipage}}            & \multicolumn{1}{c|}{\begin{minipage}[b]{0.02\columnwidth}\centering\raisebox{-.2\height}{\includegraphics[width=\linewidth]{fig/icon/annulus.pdf}}
	\end{minipage}}            &  \begin{minipage}[b]{0.02\columnwidth}\centering\raisebox{-.2\height}{\includegraphics[width=\linewidth]{fig/icon/circle.pdf}}
	\end{minipage}                                & \multicolumn{1}{c|}{\begin{minipage}[b]{0.02\columnwidth}\centering\raisebox{-.2\height}{\includegraphics[width=\linewidth]{fig/icon/annulus.pdf}}
	\end{minipage}}            & \multicolumn{1}{c|}{\begin{minipage}[b]{0.02\columnwidth}\centering\raisebox{-.2\height}{\includegraphics[width=\linewidth]{fig/icon/circle.pdf}}
	\end{minipage}}            &   \begin{minipage}[b]{0.02\columnwidth}\centering\raisebox{-.2\height}{\includegraphics[width=\linewidth]{fig/icon/semicircle.pdf}}
	\end{minipage}                               
	\\  \cline{2-13} 
   &\cite{cui2021efficient}  &\multicolumn{1}{c|}{O-E}  &  TO-UE  & \multicolumn{1}{c|}{\begin{minipage}[b]{0.02\columnwidth}\centering\raisebox{-.2\height}{\includegraphics[width=\linewidth]{fig/icon/annulus.pdf}}
	\end{minipage}}   & \multicolumn{1}{c|}{\begin{minipage}[b]{0.02\columnwidth}\centering\raisebox{-.2\height}{\includegraphics[width=\linewidth]{fig/icon/circle.pdf}}
	\end{minipage}}            &   \begin{minipage}[b]{0.02\columnwidth}\centering\raisebox{-.2\height}{\includegraphics[width=\linewidth]{fig/icon/circle.pdf}}
	\end{minipage}     & \multicolumn{1}{c|}{\begin{minipage}[b]{0.02\columnwidth}\centering\raisebox{-.2\height}{\includegraphics[width=\linewidth]{fig/icon/annulus.pdf}}
	\end{minipage}}            & \multicolumn{1}{c|}{\begin{minipage}[b]{0.02\columnwidth}\centering\raisebox{-.2\height}{\includegraphics[width=\linewidth]{fig/icon/annulus.pdf}}
	\end{minipage}}            &   \begin{minipage}[b]{0.02\columnwidth}\centering\raisebox{-.2\height}{\includegraphics[width=\linewidth]{fig/icon/circle.pdf}}
	\end{minipage}                               & \multicolumn{1}{c|}{\begin{minipage}[b]{0.02\columnwidth}\centering\raisebox{-.2\height}{\includegraphics[width=\linewidth]{fig/icon/annulus.pdf}}
	\end{minipage}}            & \multicolumn{1}{c|}{\begin{minipage}[b]{0.02\columnwidth}\centering\raisebox{-.2\height}{\includegraphics[width=\linewidth]{fig/icon/circle.pdf}}
	\end{minipage}}            &     \begin{minipage}[b]{0.02\columnwidth}\centering\raisebox{-.2\height}{\includegraphics[width=\linewidth]{fig/icon/semicircle.pdf}}
	\end{minipage}                            \\ \cline{2-13} & \cite{qiao2022lightweight}&\multicolumn{1}{c|}{O-E}  & TO-UE  & \multicolumn{1}{c|}{\begin{minipage}[b]{0.02\columnwidth}\centering\raisebox{-.2\height}{\includegraphics[width=\linewidth]{fig/icon/circle.pdf}}
	\end{minipage}}            & \multicolumn{1}{c|}{\begin{minipage}[b]{0.02\columnwidth}\centering\raisebox{-.2\height}{\includegraphics[width=\linewidth]{fig/icon/circle.pdf}}
	\end{minipage}}            & \begin{minipage}[b]{0.02\columnwidth}\centering\raisebox{-.2\height}{\includegraphics[width=\linewidth]{fig/icon/circle.pdf}}
	\end{minipage}  & \multicolumn{1}{c|}{\begin{minipage}[b]{0.02\columnwidth}\centering\raisebox{-.2\height}{\includegraphics[width=\linewidth]{fig/icon/annulus.pdf}}
	\end{minipage}}            & \multicolumn{1}{c|}{\begin{minipage}[b]{0.02\columnwidth}\centering\raisebox{-.2\height}{\includegraphics[width=\linewidth]{fig/icon/annulus.pdf}}
	\end{minipage}}            &\begin{minipage}[b]{0.02\columnwidth}\centering\raisebox{-.2\height}{\includegraphics[width=\linewidth]{fig/icon/circle.pdf}}
	\end{minipage}  & \multicolumn{1}{c|}{\begin{minipage}[b]{0.02\columnwidth}\centering\raisebox{-.2\height}{\includegraphics[width=\linewidth]{fig/icon/circle.pdf}}
	\end{minipage}} & \multicolumn{1}{c|}{\begin{minipage}[b]{0.02\columnwidth}\centering\raisebox{-.2\height}{\includegraphics[width=\linewidth]{fig/icon/circle.pdf}}
	\end{minipage}}            & \begin{minipage}[b]{0.02\columnwidth}\centering\raisebox{-.2\height}{\includegraphics[width=\linewidth]{fig/icon/semicircle.pdf}}
	\end{minipage}    \\ \cline{2-13} & \cite{ding2022edge}&\multicolumn{1}{c|}{O-E}  &  TO-UE  & \multicolumn{1}{c|}{\begin{minipage}[b]{0.02\columnwidth}\centering\raisebox{-.2\height}{\includegraphics[width=\linewidth]{fig/icon/circle.pdf}}
	\end{minipage}}            & \multicolumn{1}{c|}{\begin{minipage}[b]{0.02\columnwidth}\centering\raisebox{-.2\height}{\includegraphics[width=\linewidth]{fig/icon/circle.pdf}}
	\end{minipage}}            &  \begin{minipage}[b]{0.02\columnwidth}\centering\raisebox{-.2\height}{\includegraphics[width=\linewidth]{fig/icon/circle.pdf}}
	\end{minipage} & \multicolumn{1}{c|}{\begin{minipage}[b]{0.02\columnwidth}\centering\raisebox{-.2\height}{\includegraphics[width=\linewidth]{fig/icon/annulus.pdf}}
	\end{minipage}}            & \multicolumn{1}{c|}{\begin{minipage}[b]{0.02\columnwidth}\centering\raisebox{-.2\height}{\includegraphics[width=\linewidth]{fig/icon/annulus.pdf}}
	\end{minipage}}            &  \begin{minipage}[b]{0.02\columnwidth}\centering\raisebox{-.2\height}{\includegraphics[width=\linewidth]{fig/icon/circle.pdf}}
	\end{minipage}& \multicolumn{1}{c|}{\begin{minipage}[b]{0.02\columnwidth}\centering\raisebox{-.2\height}{\includegraphics[width=\linewidth]{fig/icon/circle.pdf}}
	\end{minipage}} & \multicolumn{1}{c|}{\begin{minipage}[b]{0.02\columnwidth}\centering\raisebox{-.2\height}{\includegraphics[width=\linewidth]{fig/icon/circle.pdf}}
	\end{minipage}}            &    \begin{minipage}[b]{0.02\columnwidth}\centering\raisebox{-.2\height}{\includegraphics[width=\linewidth]{fig/icon/semicircle.pdf}}
	\end{minipage}        \\ \cline{2-13} & \cite{mei2022blockchain}&\multicolumn{1}{c|}{O-E}  &  TO-TE  & \multicolumn{1}{c|}{\begin{minipage}[b]{0.02\columnwidth}\centering\raisebox{-.2\height}{\includegraphics[width=\linewidth]{fig/icon/circle.pdf}}
	\end{minipage}}            & \multicolumn{1}{c|}{\begin{minipage}[b]{0.02\columnwidth}\centering\raisebox{-.2\height}{\includegraphics[width=\linewidth]{fig/icon/circle.pdf}}
	\end{minipage}}            &  \begin{minipage}[b]{0.02\columnwidth}\centering\raisebox{-.2\height}{\includegraphics[width=\linewidth]{fig/icon/circle.pdf}}
	\end{minipage} & \multicolumn{1}{c|}{\begin{minipage}[b]{0.02\columnwidth}\centering\raisebox{-.2\height}{\includegraphics[width=\linewidth]{fig/icon/annulus.pdf}}
	\end{minipage}}            & \multicolumn{1}{c|}{\begin{minipage}[b]{0.02\columnwidth}\centering\raisebox{-.2\height}{\includegraphics[width=\linewidth]{fig/icon/annulus.pdf}}
	\end{minipage}}            &  \begin{minipage}[b]{0.02\columnwidth}\centering\raisebox{-.2\height}{\includegraphics[width=\linewidth]{fig/icon/circle.pdf}}
	\end{minipage}& \multicolumn{1}{c|}{\begin{minipage}[b]{0.02\columnwidth}\centering\raisebox{-.2\height}{\includegraphics[width=\linewidth]{fig/icon/annulus.pdf}}
	\end{minipage}} & \multicolumn{1}{c|}{\begin{minipage}[b]{0.02\columnwidth}\centering\raisebox{-.2\height}{\includegraphics[width=\linewidth]{fig/icon/circle.pdf}}
	\end{minipage}}            &    \begin{minipage}[b]{0.02\columnwidth}\centering\raisebox{-.2\height}{\includegraphics[width=\linewidth]{fig/icon/semicircle.pdf}}
	\end{minipage}       \\ \cline{2-13} & \cite{li2023or}&\multicolumn{1}{c|}{O-E}  &  TO-UE  & \multicolumn{1}{c|}{\begin{minipage}[b]{0.02\columnwidth}\centering\raisebox{-.2\height}{\includegraphics[width=\linewidth]{fig/icon/circle.pdf}}
	\end{minipage}}            & \multicolumn{1}{c|}{\begin{minipage}[b]{0.02\columnwidth}\centering\raisebox{-.2\height}{\includegraphics[width=\linewidth]{fig/icon/circle.pdf}}
	\end{minipage}}            &  \begin{minipage}[b]{0.02\columnwidth}\centering\raisebox{-.2\height}{\includegraphics[width=\linewidth]{fig/icon/circle.pdf}}
	\end{minipage} & \multicolumn{1}{c|}{\begin{minipage}[b]{0.02\columnwidth}\centering\raisebox{-.2\height}{\includegraphics[width=\linewidth]{fig/icon/annulus.pdf}}
	\end{minipage}}            & \multicolumn{1}{c|}{\begin{minipage}[b]{0.02\columnwidth}\centering\raisebox{-.2\height}{\includegraphics[width=\linewidth]{fig/icon/annulus.pdf}}
	\end{minipage}}            &  \begin{minipage}[b]{0.02\columnwidth}\centering\raisebox{-.2\height}{\includegraphics[width=\linewidth]{fig/icon/circle.pdf}}
	\end{minipage}& \multicolumn{1}{c|}{\begin{minipage}[b]{0.02\columnwidth}\centering\raisebox{-.2\height}{\includegraphics[width=\linewidth]{fig/icon/annulus.pdf}}
	\end{minipage}} & \multicolumn{1}{c|}{\begin{minipage}[b]{0.02\columnwidth}\centering\raisebox{-.2\height}{\includegraphics[width=\linewidth]{fig/icon/circle.pdf}}
	\end{minipage}}            &    \begin{minipage}[b]{0.02\columnwidth}\centering\raisebox{-.2\height}{\includegraphics[width=\linewidth]{fig/icon/semicircle.pdf}}
	\end{minipage} \\ \cline{2-13} & \cite{sellami2024verifiable}&\multicolumn{1}{c|}{O-E}  &  TO-UE  & \multicolumn{1}{c|}{\begin{minipage}[b]{0.02\columnwidth}\centering\raisebox{-.2\height}{\includegraphics[width=\linewidth]{fig/icon/circle.pdf}}
	\end{minipage}}            & \multicolumn{1}{c|}{\begin{minipage}[b]{0.02\columnwidth}\centering\raisebox{-.2\height}{\includegraphics[width=\linewidth]{fig/icon/circle.pdf}}
	\end{minipage}}            &  \begin{minipage}[b]{0.02\columnwidth}\centering\raisebox{-.2\height}{\includegraphics[width=\linewidth]{fig/icon/annulus.pdf}}
	\end{minipage} & \multicolumn{1}{c|}{\begin{minipage}[b]{0.02\columnwidth}\centering\raisebox{-.2\height}{\includegraphics[width=\linewidth]{fig/icon/annulus.pdf}}
	\end{minipage}}            & \multicolumn{1}{c|}{\begin{minipage}[b]{0.02\columnwidth}\centering\raisebox{-.2\height}{\includegraphics[width=\linewidth]{fig/icon/annulus.pdf}}
	\end{minipage}}            &  \begin{minipage}[b]{0.02\columnwidth}\centering\raisebox{-.2\height}{\includegraphics[width=\linewidth]{fig/icon/circle.pdf}}
	\end{minipage}& \multicolumn{1}{c|}{\begin{minipage}[b]{0.02\columnwidth}\centering\raisebox{-.2\height}{\includegraphics[width=\linewidth]{fig/icon/annulus.pdf}}
	\end{minipage}} & \multicolumn{1}{c|}{\begin{minipage}[b]{0.02\columnwidth}\centering\raisebox{-.2\height}{\includegraphics[width=\linewidth]{fig/icon/circle.pdf}}
	\end{minipage}}            &    \begin{minipage}[b]{0.02\columnwidth}\centering\raisebox{-.2\height}{\includegraphics[width=\linewidth]{fig/icon/semicircle.pdf}}
	\end{minipage} \\ \cline{2-13} & \cite{maheswari2023clustering}&\multicolumn{1}{c|}{O-E}  &  TO-UE  & \multicolumn{1}{c|}{\begin{minipage}[b]{0.02\columnwidth}\centering\raisebox{-.2\height}{\includegraphics[width=\linewidth]{fig/icon/circle.pdf}}
	\end{minipage}}            & \multicolumn{1}{c|}{\begin{minipage}[b]{0.02\columnwidth}\centering\raisebox{-.2\height}{\includegraphics[width=\linewidth]{fig/icon/circle.pdf}}
	\end{minipage}}            &  \begin{minipage}[b]{0.02\columnwidth}\centering\raisebox{-.2\height}{\includegraphics[width=\linewidth]{fig/icon/circle.pdf}}
	\end{minipage} & \multicolumn{1}{c|}{\begin{minipage}[b]{0.02\columnwidth}\centering\raisebox{-.2\height}{\includegraphics[width=\linewidth]{fig/icon/annulus.pdf}}
	\end{minipage}}            & \multicolumn{1}{c|}{\begin{minipage}[b]{0.02\columnwidth}\centering\raisebox{-.2\height}{\includegraphics[width=\linewidth]{fig/icon/annulus.pdf}}
	\end{minipage}}            &  \begin{minipage}[b]{0.02\columnwidth}\centering\raisebox{-.2\height}{\includegraphics[width=\linewidth]{fig/icon/circle.pdf}}
	\end{minipage}& \multicolumn{1}{c|}{\begin{minipage}[b]{0.02\columnwidth}\centering\raisebox{-.2\height}{\includegraphics[width=\linewidth]{fig/icon/circle.pdf}}
	\end{minipage}} & \multicolumn{1}{c|}{\begin{minipage}[b]{0.02\columnwidth}\centering\raisebox{-.2\height}{\includegraphics[width=\linewidth]{fig/icon/circle.pdf}}
	\end{minipage}}            &    \begin{minipage}[b]{0.02\columnwidth}\centering\raisebox{-.2\height}{\includegraphics[width=\linewidth]{fig/icon/semicircle.pdf}}
	\end{minipage} \\ \hline
\multirow{7}{*}{\begin{tabular}[c]{@{}c@{}}Public\\ Audit\end{tabular}}        & \cite{tong2019privacy}         & \multicolumn{1}{c|}{U-T-E}    &  TU-ST-UE  & \multicolumn{1}{c|}{\begin{minipage}[b]{0.02\columnwidth}\centering\raisebox{-.2\height}{\includegraphics[width=\linewidth]{fig/icon/circle.pdf}}
	\end{minipage}}            & \multicolumn{1}{c|}{\begin{minipage}[b]{0.02\columnwidth}\centering\raisebox{-.2\height}{\includegraphics[width=\linewidth]{fig/icon/circle.pdf}}
	\end{minipage}}            & \begin{minipage}[b]{0.02\columnwidth}\centering\raisebox{-.2\height}{\includegraphics[width=\linewidth]{fig/icon/circle.pdf}}
	\end{minipage}                                 & \multicolumn{1}{c|}{\begin{minipage}[b]{0.02\columnwidth}\centering\raisebox{-.2\height}{\includegraphics[width=\linewidth]{fig/icon/annulus.pdf}}
	\end{minipage}}            & \multicolumn{1}{c|}{\begin{minipage}[b]{0.02\columnwidth}\centering\raisebox{-.2\height}{\includegraphics[width=\linewidth]{fig/icon/circle.pdf}}
	\end{minipage}}            &   \begin{minipage}[b]{0.02\columnwidth}\centering\raisebox{-.2\height}{\includegraphics[width=\linewidth]{fig/icon/circle.pdf}}
	\end{minipage}                               & \multicolumn{1}{c|}{\begin{minipage}[b]{0.02\columnwidth}\centering\raisebox{-.2\height}{\includegraphics[width=\linewidth]{fig/icon/annulus.pdf}}
	\end{minipage}}            & \multicolumn{1}{c|}{\begin{minipage}[b]{0.02\columnwidth}\centering\raisebox{-.2\height}{\includegraphics[width=\linewidth]{fig/icon/circle.pdf}}
	\end{minipage}}            &    \begin{minipage}[b]{0.02\columnwidth}\centering\raisebox{-.2\height}{\includegraphics[width=\linewidth]{fig/icon/semicircle.pdf}}
	\end{minipage}                              \\ \cline{2-13} 
& \cite{liu2020efficient}        & \multicolumn{1}{c|}{O-T-E}        &  TO-TT-SE   & \multicolumn{1}{c|}{\begin{minipage}[b]{0.02\columnwidth}\centering\raisebox{-.2\height}{\includegraphics[width=\linewidth]{fig/icon/circle.pdf}}
	\end{minipage}}            & \multicolumn{1}{c|}{\begin{minipage}[b]{0.02\columnwidth}\centering\raisebox{-.2\height}{\includegraphics[width=\linewidth]{fig/icon/circle.pdf}}
	\end{minipage}}            &   \begin{minipage}[b]{0.02\columnwidth}\centering\raisebox{-.2\height}{\includegraphics[width=\linewidth]{fig/icon/circle.pdf}}
	\end{minipage}                               & \multicolumn{1}{c|}{\begin{minipage}[b]{0.02\columnwidth}\centering\raisebox{-.2\height}{\includegraphics[width=\linewidth]{fig/icon/circle.pdf}}
	\end{minipage}}            & \multicolumn{1}{c|}{\begin{minipage}[b]{0.02\columnwidth}\centering\raisebox{-.2\height}{\includegraphics[width=\linewidth]{fig/icon/circle.pdf}}
	\end{minipage}}            &  \begin{minipage}[b]{0.02\columnwidth}\centering\raisebox{-.2\height}{\includegraphics[width=\linewidth]{fig/icon/circle.pdf}}
	\end{minipage}                                & \multicolumn{1}{c|}{\begin{minipage}[b]{0.02\columnwidth}\centering\raisebox{-.2\height}{\includegraphics[width=\linewidth]{fig/icon/annulus.pdf}}
	\end{minipage}}            & \multicolumn{1}{c|}{\begin{minipage}[b]{0.02\columnwidth}\centering\raisebox{-.2\height}{\includegraphics[width=\linewidth]{fig/icon/annulus.pdf}}
	\end{minipage}}            &   \begin{minipage}[b]{0.02\columnwidth}\centering\raisebox{-.2\height}{\includegraphics[width=\linewidth]{fig/icon/semicircle.pdf}}
	\end{minipage}                               \\ \cline{2-13} 
   & \cite{wang2021zss}             &  \multicolumn{1}{c|}{O-T-E}     &  TO-ST-TE        & \multicolumn{1}{c|}{\begin{minipage}[b]{0.02\columnwidth}\centering\raisebox{-.2\height}{\includegraphics[width=\linewidth]{fig/icon/annulus.pdf}}
	\end{minipage} }            & \multicolumn{1}{c|}{\begin{minipage}[b]{0.02\columnwidth}\centering\raisebox{-.2\height}{\includegraphics[width=\linewidth]{fig/icon/circle.pdf}}
	\end{minipage} }            &  \begin{minipage}[b]{0.02\columnwidth}\centering\raisebox{-.2\height}{\includegraphics[width=\linewidth]{fig/icon/circle.pdf}}
	\end{minipage}                                 & \multicolumn{1}{c|}{\begin{minipage}[b]{0.02\columnwidth}\centering\raisebox{-.2\height}{\includegraphics[width=\linewidth]{fig/icon/annulus.pdf}}
	\end{minipage} }            & \multicolumn{1}{c|}{\begin{minipage}[b]{0.02\columnwidth}\centering\raisebox{-.2\height}{\includegraphics[width=\linewidth]{fig/icon/circle.pdf}}
	\end{minipage} }            &    \begin{minipage}[b]{0.02\columnwidth}\centering\raisebox{-.2\height}{\includegraphics[width=\linewidth]{fig/icon/circle.pdf}}
	\end{minipage}                               & \multicolumn{1}{c|}{\begin{minipage}[b]{0.02\columnwidth}\centering\raisebox{-.2\height}{\includegraphics[width=\linewidth]{fig/icon/circle.pdf}}
	\end{minipage} }            & \multicolumn{1}{c|}{\begin{minipage}[b]{0.02\columnwidth}\centering\raisebox{-.2\height}{\includegraphics[width=\linewidth]{fig/icon/annulus.pdf}}
	\end{minipage}}            &    \begin{minipage}[b]{0.02\columnwidth}\centering\raisebox{-.2\height}{\includegraphics[width=\linewidth]{fig/icon/semicircle.pdf}}
	\end{minipage}             \\ \cline{2-13} 
 & \cite{wang2022lightweight}     & \multicolumn{1}{c|}{O-T-E}           &  TO-ST-TE          & \multicolumn{1}{c|}{\begin{minipage}[b]{0.02\columnwidth}\centering\raisebox{-.2\height}{\includegraphics[width=\linewidth]{fig/icon/annulus.pdf}}
	\end{minipage}}            & \multicolumn{1}{c|}{\begin{minipage}[b]{0.02\columnwidth}\centering\raisebox{-.2\height}{\includegraphics[width=\linewidth]{fig/icon/circle.pdf}}
	\end{minipage}}            &    \begin{minipage}[b]{0.02\columnwidth}\centering\raisebox{-.2\height}{\includegraphics[width=\linewidth]{fig/icon/circle.pdf}}
	\end{minipage}                              & \multicolumn{1}{c|}{\begin{minipage}[b]{0.02\columnwidth}\centering\raisebox{-.2\height}{\includegraphics[width=\linewidth]{fig/icon/annulus.pdf}}
	\end{minipage}}            & \multicolumn{1}{c|}{\begin{minipage}[b]{0.02\columnwidth}\centering\raisebox{-.2\height}{\includegraphics[width=\linewidth]{fig/icon/circle.pdf}}
	\end{minipage}}            &   \begin{minipage}[b]{0.02\columnwidth}\centering\raisebox{-.2\height}{\includegraphics[width=\linewidth]{fig/icon/circle.pdf}}
	\end{minipage}                               & \multicolumn{1}{c|}{\begin{minipage}[b]{0.02\columnwidth}\centering\raisebox{-.2\height}{\includegraphics[width=\linewidth]{fig/icon/circle.pdf}}
	\end{minipage}}            & \multicolumn{1}{c|}{\begin{minipage}[b]{0.02\columnwidth}\centering\raisebox{-.2\height}{\includegraphics[width=\linewidth]{fig/icon/circle.pdf}}
	\end{minipage}}            &   \begin{minipage}[b]{0.02\columnwidth}\centering\raisebox{-.2\height}{\includegraphics[width=\linewidth]{fig/icon/semicircle.pdf}}
	\end{minipage}                                                 \\\cline{2-13} 
     & \cite{chen2021trusted}         & \multicolumn{1}{c|}{O-T-E}  &   UO-UT-TE & \multicolumn{1}{c|}{\begin{minipage}[b]{0.02\columnwidth}\centering\raisebox{-.2\height}{\includegraphics[width=\linewidth]{fig/icon/annulus.pdf}}
	\end{minipage}}            & \multicolumn{1}{c|}{\begin{minipage}[b]{0.02\columnwidth}\centering\raisebox{-.2\height}{\includegraphics[width=\linewidth]{fig/icon/circle.pdf}}
	\end{minipage}}            &   \begin{minipage}[b]{0.02\columnwidth}\centering\raisebox{-.2\height}{\includegraphics[width=\linewidth]{fig/icon/circle.pdf}}
	\end{minipage}                               & \multicolumn{1}{c|}{\begin{minipage}[b]{0.02\columnwidth}\centering\raisebox{-.2\height}{\includegraphics[width=\linewidth]{fig/icon/annulus.pdf}}
	\end{minipage}}            & \multicolumn{1}{c|}{\begin{minipage}[b]{0.02\columnwidth}\centering\raisebox{-.2\height}{\includegraphics[width=\linewidth]{fig/icon/circle.pdf}}
	\end{minipage}}            &  \begin{minipage}[b]{0.02\columnwidth}\centering\raisebox{-.2\height}{\includegraphics[width=\linewidth]{fig/icon/circle.pdf}}
	\end{minipage}                                & \multicolumn{1}{c|}{\begin{minipage}[b]{0.02\columnwidth}\centering\raisebox{-.2\height}{\includegraphics[width=\linewidth]{fig/icon/annulus.pdf}}
	\end{minipage}}            & \multicolumn{1}{c|}{\begin{minipage}[b]{0.02\columnwidth}\centering\raisebox{-.2\height}{\includegraphics[width=\linewidth]{fig/icon/annulus.pdf}}
	\end{minipage}}            &   \begin{minipage}[b]{0.02\columnwidth}\centering\raisebox{-.2\height}{\includegraphics[width=\linewidth]{fig/icon/semicircle.pdf}}
	\end{minipage}                               \\ \cline{2-13} 
                                                                               & \cite{tong2022privacy}         &    \multicolumn{1}{c|}{U-T-E}   &  TU-ST-UE  
                                            & \multicolumn{1}{c|}{\begin{minipage}[b]{0.02\columnwidth}\centering\raisebox{-.2\height}{\includegraphics[width=\linewidth]{fig/icon/circle.pdf}}
	\end{minipage}}            & \multicolumn{1}{c|}{\begin{minipage}[b]{0.02\columnwidth}\centering\raisebox{-.2\height}{\includegraphics[width=\linewidth]{fig/icon/circle.pdf}}
	\end{minipage}}            & \begin{minipage}[b]{0.02\columnwidth}\centering\raisebox{-.2\height}{\includegraphics[width=\linewidth]{fig/icon/circle.pdf}}
	\end{minipage}                                 & \multicolumn{1}{c|}{\begin{minipage}[b]{0.02\columnwidth}\centering\raisebox{-.2\height}{\includegraphics[width=\linewidth]{fig/icon/annulus.pdf}}
	\end{minipage}}            & \multicolumn{1}{c|}{\begin{minipage}[b]{0.02\columnwidth}\centering\raisebox{-.2\height}{\includegraphics[width=\linewidth]{fig/icon/circle.pdf}}
	\end{minipage}}            &   \begin{minipage}[b]{0.02\columnwidth}\centering\raisebox{-.2\height}{\includegraphics[width=\linewidth]{fig/icon/circle.pdf}}
	\end{minipage}                               & \multicolumn{1}{c|}{\begin{minipage}[b]{0.02\columnwidth}\centering\raisebox{-.2\height}{\includegraphics[width=\linewidth]{fig/icon/annulus.pdf}}
	\end{minipage}}            & \multicolumn{1}{c|}{\begin{minipage}[b]{0.02\columnwidth}\centering\raisebox{-.2\height}{\includegraphics[width=\linewidth]{fig/icon/circle.pdf}}
	\end{minipage}}            &    \begin{minipage}[b]{0.02\columnwidth}\centering\raisebox{-.2\height}{\includegraphics[width=\linewidth]{fig/icon/semicircle.pdf}}
	\end{minipage}     \\ \cline{2-13} 
                & \cite{liu2022secure}         &    \multicolumn{1}{c|}{U-T-E}   &  TU-TT-SE
                                            & \multicolumn{1}{c|}{\begin{minipage}[b]{0.02\columnwidth}\centering\raisebox{-.2\height}{\includegraphics[width=\linewidth]{fig/icon/annulus.pdf}}
	\end{minipage}}            & \multicolumn{1}{c|}{\begin{minipage}[b]{0.02\columnwidth}\centering\raisebox{-.2\height}{\includegraphics[width=\linewidth]{fig/icon/circle.pdf}}
	\end{minipage}}            &  \begin{minipage}[b]{0.02\columnwidth}\centering\raisebox{-.2\height}{\includegraphics[width=\linewidth]{fig/icon/circle.pdf}}
	\end{minipage}                                & \multicolumn{1}{c|}{\begin{minipage}[b]{0.02\columnwidth}\centering\raisebox{-.2\height}{\includegraphics[width=\linewidth]{fig/icon/annulus.pdf}}
	\end{minipage}}            & \multicolumn{1}{c|}{\begin{minipage}[b]{0.02\columnwidth}\centering\raisebox{-.2\height}{\includegraphics[width=\linewidth]{fig/icon/circle.pdf}}
	\end{minipage}}            &   \begin{minipage}[b]{0.02\columnwidth}\centering\raisebox{-.2\height}{\includegraphics[width=\linewidth]{fig/icon/circle.pdf}}
	\end{minipage}                               & \multicolumn{1}{c|}{\begin{minipage}[b]{0.02\columnwidth}\centering\raisebox{-.2\height}{\includegraphics[width=\linewidth]{fig/icon/annulus.pdf}}
	\end{minipage}}            & \multicolumn{1}{c|}{\begin{minipage}[b]{0.02\columnwidth}\centering\raisebox{-.2\height}{\includegraphics[width=\linewidth]{fig/icon/circle.pdf}}
	\end{minipage}}            &  \begin{minipage}[b]{0.02\columnwidth}\centering\raisebox{-.2\height}{\includegraphics[width=\linewidth]{fig/icon/semicircle.pdf}}
	\end{minipage}  \\ \hline
\multirow{6}{*}{\begin{tabular}[c]{@{}c@{}}Cooperative\\ Audit\end{tabular}} & \cite{alazeb2019ensuring}      &\multicolumn{1}{c|}{E-E}                                                              &   TE-TE                                                      & \multicolumn{1}{c|}{\begin{minipage}[b]{0.02\columnwidth}
		\centering
		\raisebox{-.2\height}{\includegraphics[width=\linewidth]{fig/icon/annulus.pdf}}
	\end{minipage}}            & \multicolumn{1}{c|}{\begin{minipage}[b]{0.02\columnwidth}
		\centering
		\raisebox{-.2\height}{\includegraphics[width=\linewidth]{fig/icon/circle.pdf}}
	\end{minipage}}            &  \begin{minipage}[b]{0.02\columnwidth}
		\centering
		\raisebox{-.2\height}{\includegraphics[width=\linewidth]{fig/icon/circle.pdf}}
	\end{minipage}                                & \multicolumn{1}{c|}{\begin{minipage}[b]{0.02\columnwidth}
		\centering
		\raisebox{-.2\height}{\includegraphics[width=\linewidth]{fig/icon/annulus.pdf}}
	\end{minipage}}            & \multicolumn{1}{c|}{\begin{minipage}[b]{0.02\columnwidth}
		\centering
		\raisebox{-.2\height}{\includegraphics[width=\linewidth]{fig/icon/annulus.pdf}}
	\end{minipage}}            &   \begin{minipage}[b]{0.02\columnwidth}
		\centering
		\raisebox{-.2\height}{\includegraphics[width=\linewidth]{fig/icon/circle.pdf}}
	\end{minipage}                               & \multicolumn{1}{c|}{\begin{minipage}[b]{0.02\columnwidth}
		\centering
		\raisebox{-.2\height}{\includegraphics[width=\linewidth]{fig/icon/annulus.pdf}}
	\end{minipage}}            & \multicolumn{1}{c|}{\begin{minipage}[b]{0.02\columnwidth}
		\centering
		\raisebox{-.2\height}{\includegraphics[width=\linewidth]{fig/icon/circle.pdf}}
	\end{minipage}}            &  \begin{minipage}[b]{0.02\columnwidth}
		\centering
		\raisebox{-.2\height}{\includegraphics[width=\linewidth]{fig/icon/semicircle.pdf}}
	\end{minipage}                                \\ \cline{2-13}                     &\cite{yue2020blockchain}       &   \multicolumn{1}{c|}{E-E}              &    UE-UE    & \multicolumn{1}{c|}{\begin{minipage}[b]{0.02\columnwidth}
		\centering
		\raisebox{-.2\height}{\includegraphics[width=\linewidth]{fig/icon/annulus.pdf}}
	\end{minipage}}            & \multicolumn{1}{c|}{\begin{minipage}[b]{0.02\columnwidth}
		\centering
		\raisebox{-.2\height}{\includegraphics[width=\linewidth]{fig/icon/circle.pdf}}
	\end{minipage}}            &   \begin{minipage}[b]{0.02\columnwidth}
		\centering
		\raisebox{-.2\height}{\includegraphics[width=\linewidth]{fig/icon/circle.pdf}}
	\end{minipage}     & \multicolumn{1}{c|}{\begin{minipage}[b]{0.02\columnwidth}
		\centering
		\raisebox{-.2\height}{\includegraphics[width=\linewidth]{fig/icon/annulus.pdf}}
	\end{minipage}}            & \multicolumn{1}{c|}{\begin{minipage}[b]{0.02\columnwidth}
		\centering
		\raisebox{-.2\height}{\includegraphics[width=\linewidth]{fig/icon/circle.pdf}}
	\end{minipage}}            &    \begin{minipage}[b]{0.02\columnwidth}
		\centering
		\raisebox{-.2\height}{\includegraphics[width=\linewidth]{fig/icon/circle.pdf}}
	\end{minipage}                              & \multicolumn{1}{c|}{\begin{minipage}[b]{0.02\columnwidth}
		\centering
		\raisebox{-.2\height}{\includegraphics[width=\linewidth]{fig/icon/annulus.pdf}}
	\end{minipage}}            & \multicolumn{1}{c|}{\begin{minipage}[b]{0.02\columnwidth}
		\centering
		\raisebox{-.2\height}{\includegraphics[width=\linewidth]{fig/icon/circle.pdf}}
	\end{minipage}}            &   \begin{minipage}[b]{0.02\columnwidth}
		\centering
		\raisebox{-.2\height}{\includegraphics[width=\linewidth]{fig/icon/semicircle.pdf}}
	\end{minipage}        \\ \cline{2-13} 
 & \cite{john2020binary}          &   \multicolumn{1}{c|}{E-E}                        &   TE-TE   & \multicolumn{1}{c|}{\begin{minipage}[b]{0.02\columnwidth}
		\centering
		\raisebox{-.2\height}{\includegraphics[width=\linewidth]{fig/icon/annulus.pdf}}
	\end{minipage}}            & \multicolumn{1}{c|}{\begin{minipage}[b]{0.02\columnwidth}
		\centering
		\raisebox{-.2\height}{\includegraphics[width=\linewidth]{fig/icon/circle.pdf}}
	\end{minipage}}            &   \begin{minipage}[b]{0.02\columnwidth}
		\centering
		\raisebox{-.2\height}{\includegraphics[width=\linewidth]{fig/icon/circle.pdf}}
	\end{minipage}                               & \multicolumn{1}{c|}{\begin{minipage}[b]{0.02\columnwidth}
		\centering
		\raisebox{-.2\height}{\includegraphics[width=\linewidth]{fig/icon/annulus.pdf}}
	\end{minipage}}            & \multicolumn{1}{c|}{\begin{minipage}[b]{0.02\columnwidth}
		\centering
		\raisebox{-.2\height}{\includegraphics[width=\linewidth]{fig/icon/annulus.pdf}}
	\end{minipage}}            &  \begin{minipage}[b]{0.02\columnwidth}
		\centering
		\raisebox{-.2\height}{\includegraphics[width=\linewidth]{fig/icon/circle.pdf}}
	\end{minipage}                                & \multicolumn{1}{c|}{\begin{minipage}[b]{0.02\columnwidth}
		\centering
		\raisebox{-.2\height}{\includegraphics[width=\linewidth]{fig/icon/annulus.pdf}}
	\end{minipage}}            & \multicolumn{1}{c|}{\begin{minipage}[b]{0.02\columnwidth}
		\centering
		\raisebox{-.2\height}{\includegraphics[width=\linewidth]{fig/icon/circle.pdf}}
	\end{minipage}}            &   \begin{minipage}[b]{0.02\columnwidth}
		\centering
		\raisebox{-.2\height}{\includegraphics[width=\linewidth]{fig/icon/semicircle.pdf}}
	\end{minipage}     \\ \cline{2-13} 
     & \cite{li2021cooperative}       &   \multicolumn{1}{c|}{E-E}                 &   TE-TE   & \multicolumn{1}{c|}{\begin{minipage}[b]{0.02\columnwidth}
		\centering
		\raisebox{-.2\height}{\includegraphics[width=\linewidth]{fig/icon/annulus.pdf}}
	\end{minipage}}            & \multicolumn{1}{c|}{\begin{minipage}[b]{0.02\columnwidth}
		\centering
		\raisebox{-.2\height}{\includegraphics[width=\linewidth]{fig/icon/circle.pdf}}
	\end{minipage}}            &   \begin{minipage}[b]{0.02\columnwidth}
		\centering
		\raisebox{-.2\height}{\includegraphics[width=\linewidth]{fig/icon/circle.pdf}}
	\end{minipage}                               & \multicolumn{1}{c|}{\begin{minipage}[b]{0.02\columnwidth}
		\centering
		\raisebox{-.2\height}{\includegraphics[width=\linewidth]{fig/icon/circle.pdf}}
	\end{minipage}}            & \multicolumn{1}{c|}{\begin{minipage}[b]{0.02\columnwidth}
		\centering
		\raisebox{-.2\height}{\includegraphics[width=\linewidth]{fig/icon/annulus.pdf}}
	\end{minipage}}            &   \begin{minipage}[b]{0.02\columnwidth}
		\centering
		\raisebox{-.2\height}{\includegraphics[width=\linewidth]{fig/icon/circle.pdf}}
	\end{minipage}                               & \multicolumn{1}{c|}{\begin{minipage}[b]{0.02\columnwidth}
		\centering
		\raisebox{-.2\height}{\includegraphics[width=\linewidth]{fig/icon/annulus.pdf}}
	\end{minipage}}            & \multicolumn{1}{c|}{\begin{minipage}[b]{0.02\columnwidth}
		\centering
		\raisebox{-.2\height}{\includegraphics[width=\linewidth]{fig/icon/circle.pdf}}
	\end{minipage}}            &     \begin{minipage}[b]{0.02\columnwidth}
		\centering
		\raisebox{-.2\height}{\includegraphics[width=\linewidth]{fig/icon/semicircle.pdf}}
	\end{minipage}     \\ \cline{2-13} & \cite{duan2022edge}            &    \multicolumn{1}{c|}{E-E}                                                      &    UE-UE                          & \multicolumn{1}{c|}{\begin{minipage}[b]{0.02\columnwidth}
		\centering
		\raisebox{-.2\height}{\includegraphics[width=\linewidth]{fig/icon/annulus.pdf}}
	\end{minipage}}            & \multicolumn{1}{c|}{\begin{minipage}[b]{0.02\columnwidth}
		\centering
		\raisebox{-.2\height}{\includegraphics[width=\linewidth]{fig/icon/circle.pdf}}
	\end{minipage}}            & \begin{minipage}[b]{0.02\columnwidth}
		\centering
		\raisebox{-.2\height}{\includegraphics[width=\linewidth]{fig/icon/circle.pdf}}
	\end{minipage}                              & \multicolumn{1}{c|}{\begin{minipage}[b]{0.02\columnwidth}
		\centering
		\raisebox{-.2\height}{\includegraphics[width=\linewidth]{fig/icon/annulus.pdf}}
	\end{minipage}}            & \multicolumn{1}{c|}{\begin{minipage}[b]{0.02\columnwidth}
		\centering
		\raisebox{-.2\height}{\includegraphics[width=\linewidth]{fig/icon/annulus.pdf}}
	\end{minipage}}            &  \begin{minipage}[b]{0.02\columnwidth}
		\centering
		\raisebox{-.2\height}{\includegraphics[width=\linewidth]{fig/icon/circle.pdf}}
	\end{minipage}                                & \multicolumn{1}{c|}{\begin{minipage}[b]{0.02\columnwidth}
		\centering
		\raisebox{-.2\height}{\includegraphics[width=\linewidth]{fig/icon/annulus.pdf}}
	\end{minipage}}            & \multicolumn{1}{c|}{\begin{minipage}[b]{0.02\columnwidth}
		\centering
		\raisebox{-.2\height}{\includegraphics[width=\linewidth]{fig/icon/circle.pdf}}
	\end{minipage}}            &   \begin{minipage}[b]{0.02\columnwidth}
		\centering
		\raisebox{-.2\height}{\includegraphics[width=\linewidth]{fig/icon/semicircle.pdf}}
	\end{minipage}    \\ \cline{2-13} 
                                                                               & \cite{li2022edgewatch}         &  \multicolumn{1}{c|}{E-E}   &  UE-UE
                                    & \multicolumn{1}{c|}{\begin{minipage}[b]{0.02\columnwidth}
		\centering
		\raisebox{-.2\height}{\includegraphics[width=\linewidth]{fig/icon/annulus.pdf}}
	\end{minipage}}            & \multicolumn{1}{c|}{\begin{minipage}[b]{0.02\columnwidth}
		\centering
		\raisebox{-.2\height}{\includegraphics[width=\linewidth]{fig/icon/circle.pdf}}
	\end{minipage}}            &  \begin{minipage}[b]{0.02\columnwidth}
		\centering
		\raisebox{-.2\height}{\includegraphics[width=\linewidth]{fig/icon/circle.pdf}}
	\end{minipage}                                & \multicolumn{1}{c|}{\begin{minipage}[b]{0.02\columnwidth}
		\centering
		\raisebox{-.2\height}{\includegraphics[width=\linewidth]{fig/icon/annulus.pdf}}
	\end{minipage}}            & \multicolumn{1}{c|}{\begin{minipage}[b]{0.02\columnwidth}
		\centering
		\raisebox{-.2\height}{\includegraphics[width=\linewidth]{fig/icon/annulus.pdf}}
	\end{minipage}}            &    \begin{minipage}[b]{0.02\columnwidth}
		\centering
		\raisebox{-.2\height}{\includegraphics[width=\linewidth]{fig/icon/circle.pdf}}
	\end{minipage}                              & \multicolumn{1}{c|}{\begin{minipage}[b]{0.02\columnwidth}
		\centering
		\raisebox{-.2\height}{\includegraphics[width=\linewidth]{fig/icon/annulus.pdf}}
	\end{minipage}}            & \multicolumn{1}{c|}{\begin{minipage}[b]{0.02\columnwidth}
		\centering
		\raisebox{-.2\height}{\includegraphics[width=\linewidth]{fig/icon/circle.pdf}}
	\end{minipage}}            &    \begin{minipage}[b]{0.02\columnwidth}
		\centering
		\raisebox{-.2\height}{\includegraphics[width=\linewidth]{fig/icon/semicircle.pdf}}
	\end{minipage}           \\ \cline{2-13} & \cite{zhao2023data}&\multicolumn{1}{c|}{E-E}  &  UE-UE  & \multicolumn{1}{c|}{\begin{minipage}[b]{0.02\columnwidth}\centering\raisebox{-.2\height}{\includegraphics[width=\linewidth]{fig/icon/circle.pdf}}
	\end{minipage}}            & \multicolumn{1}{c|}{\begin{minipage}[b]{0.02\columnwidth}\centering\raisebox{-.2\height}{\includegraphics[width=\linewidth]{fig/icon/circle.pdf}}
	\end{minipage}}            &  \begin{minipage}[b]{0.02\columnwidth}\centering\raisebox{-.2\height}{\includegraphics[width=\linewidth]{fig/icon/circle.pdf}}
	\end{minipage} & \multicolumn{1}{c|}{\begin{minipage}[b]{0.02\columnwidth}\centering\raisebox{-.2\height}{\includegraphics[width=\linewidth]{fig/icon/annulus.pdf}}
	\end{minipage}}            & \multicolumn{1}{c|}{\begin{minipage}[b]{0.02\columnwidth}\centering\raisebox{-.2\height}{\includegraphics[width=\linewidth]{fig/icon/circle.pdf}}
	\end{minipage}}            &  \begin{minipage}[b]{0.02\columnwidth}\centering\raisebox{-.2\height}{\includegraphics[width=\linewidth]{fig/icon/circle.pdf}}
	\end{minipage}& \multicolumn{1}{c|}{\begin{minipage}[b]{0.02\columnwidth}\centering\raisebox{-.2\height}{\includegraphics[width=\linewidth]{fig/icon/annulus.pdf}}
	\end{minipage}} & \multicolumn{1}{c|}{\begin{minipage}[b]{0.02\columnwidth}\centering\raisebox{-.2\height}{\includegraphics[width=\linewidth]{fig/icon/circle.pdf}}
	\end{minipage}}            &    \begin{minipage}[b]{0.02\columnwidth}\centering\raisebox{-.2\height}{\includegraphics[width=\linewidth]{fig/icon/semicircle.pdf}}
	\end{minipage}                \\ \hline\hline
\end{tabular}
      \begin{tablenotes}
		\item 1: O-E (DO and ENs); U-T-E (DU, TPA, and ENs); O-T-E (DO, TPA, and ENs); E-E (among ENs).
		\item 2: TO-UE (trusted DO and untrusted ENs); TO-TE (trusted DO and ENs); TU-ST-UE (trusted DU, semi-trusted TPA and untrusted ENs); TO-TT-SE (trusted DO and TPA, semi-trusted ENs); TO-ST-TE (trusted DO and ENs, and semi-trusted TPA); UO-UT-TE (untrusted DO and TPA, and trusted ENs); TU-TT-SE (trusted DU and TPA, and semi-trusted ENs); TE-TE (trusted ENs); UE-UE (untrusted ENs). [Please note that semi-trusted denotes honest-but-curious.]
		\item 3: \begin{minipage}[b]{0.02\columnwidth}\centering\raisebox{-.2\height}{\includegraphics[width=\linewidth]{fig/icon/circle.pdf}}
	\end{minipage} (support); \begin{minipage}[b]{0.02\columnwidth}\centering\raisebox{-.2\height}{\includegraphics[width=\linewidth]{fig/icon/semicircle.pdf}}
	\end{minipage} (uncertainty); \begin{minipage}[b]{0.02\columnwidth}\centering\raisebox{-.2\height}{\includegraphics[width=\linewidth]{fig/icon/annulus.pdf}}
	\end{minipage} (non-support) for efficiency (Effi.), security (Secur.), and functionality (Funct.) related indicators.
     \end{tablenotes}
\end{threeparttable}
}
\end{table*}

\begin{itemize}
   \item[2015:]
    \emph{The emergence of edge computing}. The roots of edge computing reach back around 2015, while it is also well-known as fog computing~\cite{yi2015survey} or cloudlet computing~\cite{babar2021cloudlet}. The aim is to explore the feasibility of performing computations on edge nodes through which network traffic is directed.
   \item[2019:]
   \emph{The emergence of EDIV problems}. Edge data integrity was not valued much before 2019, until Tong \emph{et. al.}~\cite{tong2019privacy} published the first work on the EDIV problem from the data users' perspective.
   \item[2020:]
   \emph{The focus on traditional approaches}. In 2020, lots of related work was proposed, but the trend is to extend traditional CDIV approaches to the edge domain without identifying the unique characteristics of edge. As we discussed in Section~\ref{subsec:Edge Data Integrity Versus Cloud Data Integrity}, there are several essential discrepancies between edge and cloud in integrity inspection.
\item[2021:]
   \emph{The focus shifted to trusted improvement}. In 2021, EDIV solutions generalized and extended the CDIV solutions space to enable better performance. This year, a significant portion of the research efforts has been dedicated to enhancing trust, especially considering that trust issues in edge are more critical compared to those in cloud. Instead of being limited to traditional approaches, EDIV solutions have generated their own specific and clear development routes.
\item[2022:]
   \emph{The focus tied to functional diversification}. In 2022, researchers began designing a more general EDIV strategy with a variety of functions. For example, they have developed integrity verification approaches in edge computing environments with data recovery and data dynamic support.
\item[2023:]
   \emph{The focus changed to scenario complexity}. From 2023 to the present, diverse EDIV scenarios have emerged, including (1) cases with multiple data owners and multiple edge nodes, where each data owner caches the same data replica on different edge nodes; and (2) cases with multiple data owners, multiple edge nodes, and multiple data replicas, where each data owner caches different data replicas on different edge nodes.
\end{itemize}

\par
\begin{figure}[!tp]
    \centering
    \includegraphics[width=0.9\linewidth]{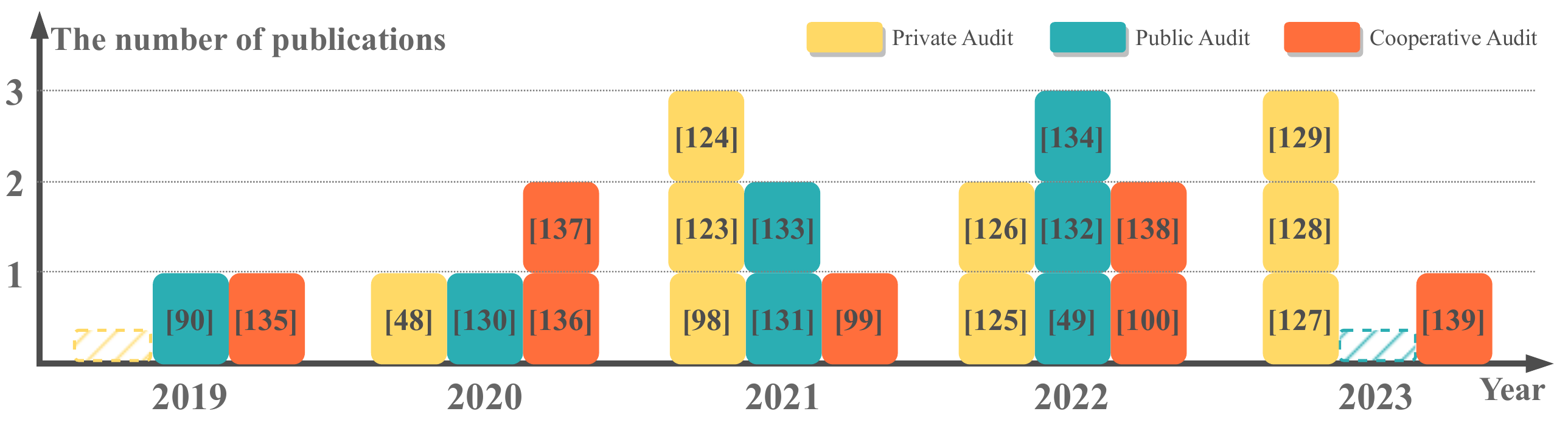}
    \caption{Graphical analysis of published papers}
    \label{fig:Graphical analysis}
    \vspace{-1 em}
\end{figure}

 Traditionally, the design philosophy of CDIV schemes heavily relies on either \textbf{provable data possession (PDP)}~\cite{ateniese2007provable} or \textbf{proof of retrievability~(POR)}~\cite{juels2007pors}. In brief, PDP schemes are probabilistic, since they employ random block sampling for verification rather than considering the entire data replica. Specifically, the original data are preprocessed to generate metadata that is stored with original data and is adopted later to verify the integrity of the cached data replica. Although this type of scheme can detect data corruption, it lacks the capability to recover it. Another well-known strategy, POR, overcomes this drawback by facilitating data recovery through redundant encoding of data. Technically, these two most-commonly used CDIV schemes are interchangeable. In fact, most existing EDIV approaches are variants of PDP or POR. To date, 23 papers on this topic have been published, as illustrated in Fig.~\ref{fig:Graphical analysis}, including 9 in private audit, 7 in public audit, and 7 in cooperative audit. Next, we review them in detail.

\subsection{Private Audit}
\label{subsec:Private Audit Papers}
Some studies concentrate on private audit, in which the data owner/user is responsible for integrity verification, with no need for TPA involvement in the whole verification process, comb-outing privacy leakage issues brought by TPA. However, the fairness issue occurs accordingly, since neither data owners/users nor edge nodes are suitable to conduct proof verification due to trust concerns~\cite{hao2020outsourced}. From a statistical perspective, private audit is the most popular scheme in academia at present. Next, we go over related works in depth.

\begin{table*}[]
\small
\caption{Summary of Recent Advances in EDIV Solutions for Private Audit}
\label{tab:Pros and Cons of Private Audit}
\setlength{\tabcolsep}{2mm}{
\begin{tabular}{m{0.5cm}<{\centering}|m{1cm}<{\centering}|m{0.5cm}<{\centering}|m{5cm}<{\centering}|m{5cm}<{\centering}|m{2.5cm}<{\centering}}
\hline \hline
\textbf{Cat.} & \textbf{Ref.} & \textbf{Year} & \textbf{Contributions} & \textbf{Limitations} & \textbf{Evaluation Metrics \& Values$^1$}  \\ \hline
\multirow{32}{*}{\rotatebox{90}{Private Audit}}
& Li \emph{et al.} \ \ \cite{li2020auditing}  & 2020        &  
\begin{itemize}[leftmargin=*]
\item[—]
    It proposes a novel data structure named variable Merkle hash tree.
  \item[—]
    It reduces verification complexity via sampling technology.
  \item[—]
    It defends against replay and forgery attacks.
\vspace{-1 em}
\end{itemize}
& 
\begin{itemize}[leftmargin=*]
  \item[—]
    It offers a probabilistic integrity guarantee, inducing some unpredictable consequences brought by undetected corruption.
  \item[—]
    It does not consider the trust issue of data owners, as well as the data dynamics and recovery issues.
  \item[—]
    It has high traffic over backhaul networks.
\vspace{-1 em}
\end{itemize} & High Computation Efficiency, High Accuracy
\\ \cline{2-6} & Li \emph{et al.}\ \ \cite{li2021inspecting} & 2021 &
\begin{itemize}[leftmargin=*]
  \item[—]
    It provides a deterministic integrity guarantee.
  \item[—]
    It supports batch verification so that efficiency could be improved.
\vspace{-1 em}
\end{itemize}   &  
\begin{itemize}[leftmargin=*]
  \item[—]
    It fails to take the security of data owners into account.
  \item[—]
    It can not repair corrupted data replicas or work well in data dynamic scenarios.
\vspace{-1 em}
\end{itemize} & High Computation Efficiency, High Accuracy
\\ \cline{2-6} & Cui \emph{et al.} \cite{cui2021efficient} & 2021       &  \begin{itemize}[leftmargin=*]
  \item[—]
    It inspects data integrity at a block level.
  \item[—]
    It aims at designing a low computation overhead solution.
\vspace{-1 em}
\end{itemize}           
& \begin{itemize}[leftmargin=*]
  \item[—]
    It has some security issues like spoofing attacks and forgery attacks.
  \item[—]
    It can not repair corrupted data and does not consider data dynamics.
\vspace{-1 em}
\end{itemize}  & High Computation Efficiency, High Accuracy   \\ \cline{2-6} & Qiao \emph{et al.} \cite{qiao2022lightweight} & 2021 &      \begin{itemize}[leftmargin=*]
  \item[—]
    It supports batch auditing and provable dynamic update.
\vspace{-1 em}
\end{itemize}                                           &  \begin{itemize}[leftmargin=*]
  \item[—]
    It can not ensure fairness or support data recovery.
\vspace{-1 em}
\end{itemize}         & High Computation Efficiency                                   \\ \cline{2-6} & Ding \emph{et al.} \cite{ding2022edge} & 2022 & \begin{itemize}[leftmargin=*]
  \item[—]
    It supports batch verification.
  \item[—]
    It proposes a new data structure named index-single linked table to support data dynamics including insertion, deletion, and modification.
\vspace{-1 em}
\end{itemize}                                               &   \begin{itemize}[leftmargin=*]
  \item[—]
    It is incapable of handling various security concerns such as forgery attacks.
  \item[—]
    It can not repair corrupted data replicas.
    \vspace{-1 em}
\end{itemize} & High Computation Efficiency \\ \cline{2-6} & Mei \emph{et al.} \cite{mei2022blockchain} & 2022 & \begin{itemize}[leftmargin=*]
  \item[—]
    It designs a blockchain-enabled privacy-preserving authentication scheme for transportation cyber-physical systems.
  \item[—]
    It supports identity privacy protection and batch verification, and simplifies key management.
\vspace{-1 em}
\end{itemize}                                               &   \begin{itemize}[leftmargin=*]
  \item[—]
    It does not consider the additional computation and communication overhead caused by blockchain.
  \item[—]
    It does not support data recovery and dynamic updates.
\vspace{-1 em}
\end{itemize} & High Computation \& Communication Efficiency \\ \cline{2-6} & Li \emph{et al.} \cite{li2023or} & 2023 & \begin{itemize}[leftmargin=*]
  \item[—]
    It explores a new scenario where different edge nodes cache different data replicas.
  \item[—]
    It proposes a per-edge one-round solution that verifies the integrity of multiple data replicas simultaneously.
\vspace{-2 em}
\end{itemize}                                               &   \begin{itemize}[leftmargin=*]
  \item[—]
    It can not handle data dynamic issues.
  \item[—]
    It may cause potential security risks, such as spoofing attacks.
\vspace{-2 em}
\end{itemize} & High Computation \& Communication Efficiency, High Accuracy  \\ \hline \hline
\end{tabular}
}
\end{table*}

\begin{table*}[]
\setcounter{table}{4}
\small
\caption{Summary of Recent Advances in EDIV Solutions for Private Audit (con.)}
\label{tab:Pros and Cons of Private Audit}
\setlength{\tabcolsep}{2mm}{
\begin{threeparttable}
\begin{tabular}{m{0.5cm}<{\centering}|m{1cm}<{\centering}|m{0.5cm}<{\centering}|m{5cm}<{\centering}|m{5cm}<{\centering}|m{2.5cm}<{\centering}}
\hline \hline
\textbf{Cat.} & \textbf{Ref.} & \textbf{Year} & \textbf{Contributions} & \textbf{Limitations} & \textbf{Evaluation Metrics \& Values$^1$}  \\ \hline
\multirow{7}{*}{\rotatebox{90}{Private Audit}}
& Sellami \emph{et al.} \cite{sellami2024verifiable} & 2023 & \begin{itemize}[leftmargin=*]
  \item[—]
    It supports data modification by distributively identifying the data owners and delegating their signatures to other entities.
  \item[—]
    It resists forgery attacks and supports data dynamic operation.
\vspace{-1 em}
\end{itemize}                                              &   \begin{itemize}[leftmargin=*]
  \item[—]
    It can not repair corrupted data replicas.
  \item[—]
    It does not support stateless verification, as data owners need to cache a signature for the entire data in each edge node.
\vspace{-1 em}
\end{itemize} & High Computation Efficiency \\ \cline{2-6} & Mahesw- ari \emph{et al.} \cite{maheswari2023clustering} & 2023 & \begin{itemize}[leftmargin=*]
  \item[—]
    It can verify the integrity of multiple data blocks within multiple data replicas at the same time.
  \item[—]
    It supports that the data owner inspects the integrity of its own replica.
\vspace{-2 em}
\end{itemize}   &   \begin{itemize}[leftmargin=*]
  \item[—]
    It does not consider data updates.
  \item[—]
    It may bring some security risks, e.g., spoofing attacks.
    \vspace{-2 em}
\end{itemize} & High Computation Efficiency \\\hline \hline
\end{tabular}
\begin{tablenotes}
		\item 1: ``High'' indicates that this approach outperforms the corresponding comparison method(s) on average for this evaluation metric. Please refer to the corresponding papers for more experimental details.
\end{tablenotes}
\end{threeparttable}
}
\end{table*}

\begin{table*}[]
\small
\caption{Summary of Recent Advances in EDIV Solutions for Public Audit}
\label{tab:Pros and Cons of Public Audit}
\setlength{\tabcolsep}{2mm}{
\begin{threeparttable}
\begin{tabular}{m{0.5cm}<{\centering}|m{1cm}<{\centering}|m{0.5cm}<{\centering}|m{5cm}<{\centering}|m{5cm}<{\centering}|m{2.5cm}<{\centering}}
\hline \hline
\textbf{Cat.} & \textbf{Ref.} & \textbf{Year} & \textbf{Contributions} & \textbf{Limitations} & \textbf{Evaluation Metrics \& Values$^1$}                        \\ \hline
\multirow{32}{*}{\rotatebox{90}{Public Audit}}
& Tong \emph{et al.} \cite{tong2019privacy} & 2019        &  \begin{itemize}[leftmargin=*]
  \item[—]
    It can verify data integrity on the edge nodes without downloading the data from them.
  \item[—]
    Both the pre-download strategy of edge nodes and the query pattern of data owners are preserved against TPA.
\vspace{-1 em}
\end{itemize}        
&         \begin{itemize}[leftmargin=*]
  \item[—]
    It is hard to ensure that TPA is totally trustworthy.
  \item[—]
    The study mainly focuses on privacy preservation and fails to tackle other unique challenges in edge. 
\item[—]
      It is a variant of the PDP scheme that limits verification efficiency.
\vspace{-1 em}
\end{itemize} & High Computation \& Communication Efficiency    \\ \cline{2-6}  & Liu \emph{et al.} \cite{liu2020efficient} & 2020       &  \begin{itemize}[leftmargin=*]
  \item[—]
    It considers data recovery by using one-way linked information tables.
  \item[—]
    It supports batch verification and thus verification efficiency can be improved.
\vspace{-1 em}
\end{itemize} &   \begin{itemize}[leftmargin=*]
  \item[—]
    It relies on an unrealistic assumption, i.e., TPA is totally trustworthy.
  \item[—]
    It barely considers possible attacks.
\vspace{-1 em}
\end{itemize}     & High Computation \& Communication Efficiency                         \\ \cline{2-6}  & Wang \emph{et al.} \cite{wang2021zss} & 2021            &    \begin{itemize}[leftmargin=*]
  \item[—]
    It considers multiple scenarios including single edge, multiple edges, and a joint of multiple edges and the cloud.
  \item[—]
    The proposed approach is privacy-preserving.
\vspace{-1 em}
\end{itemize}   &   \begin{itemize}[leftmargin=*]
  \item[—]
    It can not guarantee that TPA can be trusted.
  \item[—]
    It does not support batch verification, data recovery and dynamics.
\vspace{-1 em}
\end{itemize}   & High Computation Efficiency  
\\ \cline{2-6} 
 & Chen \emph{et al.} \cite{chen2021trusted} & 2021         &  \begin{itemize}[leftmargin=*]
  \item[—]
    It crowdsources auditing tasks to multiple auditors to solve untrusted TPA issues by using blockchain.
  \item[—]
    It proposes an unbiased selection algorithm to select TPA from the auditor committee and designs an incentive mechanism to force TPA to act honestly. 
\vspace{-1 em}
\end{itemize}                                                           & \begin{itemize}[leftmargin=*]
  \item[—]
    It does not consider data recovery and dynamics.
  \item[—]
    It does not consider protecting the privacy of data owners.
  \item[—]
    It fails to tackle possible attacks, such as collusion attacks.
\vspace{-1 em}
\end{itemize}    & High Gas Utilization            \\ \cline{2-6}    & Wang \emph{et al.} \cite{wang2022lightweight}  & 2022   &    \begin{itemize}[leftmargin=*]
  \item[—]
    It is an extension of \cite{wang2021zss}, which has the same merits as it. They further propose an optimization strategy based on a matrix index to support data dynamics.
  \item[—]
    It adopts a novel integrity-proof generation method by using an algebraic signature.
\vspace{-1 em}
\end{itemize}                                                        &   \begin{itemize}[leftmargin=*]
  \item[—]
    It has the same limitations as \cite{wang2021zss}, except for data dynamics support.
\end{itemize}            & High Computation Efficiency                             \\ \cline{2-6}  & Tong \emph{et al.} \cite{tong2022privacy}  & 2022       &   \begin{itemize}[leftmargin=*]
  \item[—]
    It is an extension of~\cite{tong2019privacy}, in which caching strategy optimization problems are investigated to store verification tags on edge nodes for communication cost reduction.
\vspace{-1 em}
\end{itemize}                                                        &  \begin{itemize}[leftmargin=*]
  \item[—]
    It has the same limitations as~\cite{tong2019privacy}.
\vspace{-1 em}
\end{itemize}  & High Computation \& Communication Efficiency \\ \cline{2-6} & Liu \emph{et al.} \cite{liu2022secure} & 2022 &   \begin{itemize}[leftmargin=*]
  \item[—]
    The proposed scheme can provide the property of key exposure resistance in auditing.
  \item[—]
    It provides privacy-preserving property.
\vspace{-1 em}
\end{itemize}            &  \begin{itemize}[leftmargin=*]
  \item[—]
    It does not support batch verification, data recovery and dynamics.
  \item[—]
    TPA may be malicious during EDIV.
\end{itemize}     & High Computation Efficiency                         \\ \hline \hline
\end{tabular}
\begin{tablenotes}
		\item 1: ``High'' indicates that this approach outperforms the corresponding comparison method(s) on average for this evaluation metric. Please refer to the corresponding papers for more experimental details.
\end{tablenotes}
\end{threeparttable} 
}
\end{table*}

\par Li \emph{et al.}~\cite{li2020auditing} (2020) propose a lightweight sampling-based probabilistic approach, namely EDI-V, aiming to audit the integrity of multiple data replicas cached on a large scale of edge nodes. Meanwhile, they develop a new data structure, variable Merkle hash tree, to facilitate audit accuracy by maintaining sampling uniformity. From the security view, EDI-V is able to defend against replay and forge attacks, while it just ensures a probabilistic integrity guarantee, which incurs security risks resulting in undetected corruption. Besides, it does not offer support to data dynamics and recovery. Moreover, the characteristic of the non-support of batch verification further constrains its practicability in large-scale edge systems. To achieve efficiency improvement, they have gone one step further to develop a deterministic EDIV approach named EDI-S~\cite{li2021inspecting} (2021) in order to support batch verification. However, EDI-S still can not fulfill corrupted data recovery or seamlessly extend to inspect dynamic data. Then, Cui \emph{et al.}~\cite{cui2021efficient} (2021) exploit a PDP-based EDIV framework named ICL-EDI by using homomorphic tags, aiming to design a low-computation solution. Like the above-described approaches, it does not support data recovery and dynamics and is easy to be damaged by various attacks.

\par Similarly, Qiao \emph{et al.}~\cite{qiao2022lightweight} (2021) develop a lightweight auditing scheme, namely EDI-SA, inspired by the shuffle algorithm and the bucket sorting algorithm. EDI-SA involves an improved sampling strategy to randomly choose data blocks to be verified. Based on algebraic signature~\cite{hevia2002provable}, EDI-SA achieves low computation overhead and supports both batch auditing and provable dynamic update. However, it does not ensure fairness, similar drawbacks marked in other private audit-based approaches. Afterward, Ding \emph{et al.}~\cite{ding2022edge} (2022) present EDI-DA, an integrity batch verification scheme. They also design a new data structure called index-single linked table to support data dynamic operation, which improves update efficiency and the practicability of the approach. On the downside, it can not carry out data recovery and still faces some security problems like forgery attacks. Mei \emph{et al.}~\cite{mei2022blockchain} (2022) propose a blockchain-based privacy-preserving authentication scheme tailored for transportation cyber-physical systems with cloud-edge collaboration. It offers unconditional anonymity and batch verification while simplifying key management through elliptic curve. Nevertheless, the integration of blockchain introduces potential computation and communication overhead, and the approach lacks support for data recovery and dynamic updates.

\par Diving into more intricate use cases, Li \emph{et al.}~\cite{li2023or} (2023) explore the scenario where distinct edge nodes cache various data replicas and try to inspect them simultaneously. Nevertheless, they fail to address data dynamic issues and potential security risks. Considering data dynamic operation, Sellami \emph{et al.}~\cite{sellami2024verifiable} (2023) design an EDIV scheme allowing data owners to keep data verifiability even when delegating data modification to edge nodes. In addition to the lack of corruption repair, it does not support stateless verification. From a different perspective, Maheswari \emph{et al.}~\cite{maheswari2023clustering} (2023) enable data owners to inspect the integrity of their own replicas, but failing to consider data updates and possible security issues, e.g., spoofing attacks.

\subsection{Public Audit}
\label{subsec:Public Audit Papers}
In some cases, it is not practically feasible for the data owner/user to remain online all the time for EDIV~\cite{zhang2019dopiv}. Hence, the data owner/user could delegate the responsibility of integrity verification to TPA to liberate itself from this heavy computing task, which derives public audit. While this approach naturally achieves fairness, it raises privacy concerns such as data leakage and user anonymity.

\par The first EDIV-related paper was published by Tong \emph{et al.}~\cite{tong2019privacy} (2019), in which they propose two integrity-checking protocols entitled ICE-basic and ICE-batch based on TPA without privacy violation. ICE-basic and ICE-batch are developed for the cases where data users inspect data integrity on a single edge node and multiple edge nodes, respectively. Even if ICE-batch supports batch verification, the verification efficiency is limited significantly due to that the proposed scheme is a variant of PDP that has been validated as not efficient enough for EDIV. Besides, they assume that TPA is fully trusted, which is hard to ensure in practice. Very recently, the same authors extended this paper in~\cite{tong2022privacy} (2022), where they try to design an effective tag cache strategy to reduce verification communication costs. Neither papers consider data dynamics and recovery. Additionally, Liu \emph{et al.}~\cite{liu2020efficient} (2020) focus on a more specific integrity verification scenario-enterprise multimedia cached on edge nodes and design an integrity auditing scheme by using homomorphic authenticator~\cite{fiore2016multi} in order to enhance computation efficiency. Meanwhile, they employ one-way linked information tables to achieve data recovery in a highly efficient manner. The advantage of it is that batch verification and data recovery are considered and handled well, and yet they barely investigate associative security, privacy, and data dynamic issues.

\par Furthermore, Wang \emph{et al.}~\cite{wang2021zss} (2021) exploit a ZSS signature~\cite{zhang2004efficient} based EDIV scheme named ZSDIVMEC with the TPA engagement, which supports privacy protection and data dynamics. They take full consideration of three usages including a single edge node, multiple edge nodes, and a joint of multiple edge nodes and a central cloud. However, batch verification is neglected, which may limit its efficiency. Furthermore, data recoverability is not discussed so that further work is needed for practicability enhancement. The same research team refines this publication and yields~\cite{wang2022lightweight} (2022). In the newest paper, they adopt algebraic signature~\cite{kipnis1998cryptanalysis} to design a lightweight EDIV framework and simultaneously design an optimized strategy for the support of data dynamics. Because it is an extension of their previous one~\cite{wang2021zss}, they have the same advantages and disadvantages, except better-supporting data dynamics, making it more practical in reality. Recently, Chen \emph{et al.}~\cite{chen2021trusted} (2021) point out that edge environments need a different trust model compared with cloud computing paradigms, as edge storage is more decentralized and thus more vulnerable to various security risks. Consequently, they devise a blockchain‐based intelligent crowdsourcing audit scheme named Crowdauditing to improve the credibility of TPAs. It totally changes the verification scheme by using an auditor committee rather than fully relying on a single TPA to achieve integrity inspection. Smart contract technology is used to collaborate with each party ensuring the reliability of audit results. Furthermore, an incentive mechanism is carefully constructed to drive auditors providing honest audit results for rewards maximum. Due to the adoption of blockchain and smart contracts, TPA-trusted issues can be well resolved, compared with other public audit schemes. However, it pays less attention to verification efficiency improvement, privacy protection, and data recovery and dynamics.

\par Recently, Liu \emph{et al.}~\cite{liu2022secure} (2022) design a EDIV scheme based on bilinear pairing~\cite{zhang2004efficient} and certificateless cryptography~\cite{al2003certificateless}. The scheme provides the property of key exposure resistance in storage auditing and supports privacy-preserving. However, the adoption of TPA is not convincing in terms of the reliability of the verification result. Besides, verification efficiency and other additional features like data recovery and dynamics support are not rigorously considered as well.

\begin{table*}[]
\small
\caption{Summary of Recent Advances in EDIV Solutions for Cooperative Audit}
\label{tab:Pros and Cons of Cooperative Audit}
\setlength{\tabcolsep}{2mm}{
\begin{threeparttable}
\begin{tabular}{m{0.5cm}<{\centering}|m{1cm}<{\centering}|m{0.5cm}<{\centering}|m{5cm}<{\centering}|m{5cm}<{\centering}|m{2.5cm}<{\centering}}
\hline \hline
\textbf{Cat.} & \textbf{Ref.} & \textbf{Year} & \textbf{Contributions} & \textbf{Limitations} & \textbf{Evaluation Metrics \& Values$^1$}                         \\ \hline
\multirow{30}{*}{\rotatebox{90}{Cooperative Audit}} & Alazeb \emph{et al.} \cite{alazeb2019ensuring} & 2019      &  \begin{itemize}[leftmargin=*]
  \item[—]
    It employs rule-based intrusion detection methods to find malicious access.
\vspace{-1 em}
\end{itemize}                                                         &    \begin{itemize}[leftmargin=*]
  \item[—]
    It does not include experimental evaluations.
  \item[—]
    It does not support batch verification, data recovery, fairness, and data dynamics.
\vspace{-1 em}
\end{itemize}      & Not Applicable                                                 \\ \cline{2-6} 
 & Yue \emph{et al.} \cite{yue2020blockchain} & 2020       &  \begin{itemize}[leftmargin=*]
  \item[—]
    It eliminates TPA by using blockchain to increase trust.
  \item[—]
    It proposes a sampling strategy to reduce verification overhead, especially for the large data replica.
\vspace{-1 em}
\end{itemize}  &  \begin{itemize}[leftmargin=*]
  \item[—]
    It has a high communication overhead incurred by blockchain.
  \item[—]
    It just considers the scenario that involves one data replica with multiple shards.
\vspace{-1 em}
\end{itemize}     & High Computation Efficiency, High Gas Utilization                                                     \\ \cline{2-6} & John \emph{et al.} \cite{john2020binary} & 2020         &  \begin{itemize}[leftmargin=*]
  \item[—]
    It explores several machine learning-based classifiers to check the integrity of electrocardiogram data.
  \item[—]
    It conducts extensive experiments to show the corruption detection performance produced by different machine learning algorithms.
\vspace{-1 em}
\end{itemize}                                                         &   \begin{itemize}[leftmargin=*]
  \item[—]
    It does not support batch verification, data recovery, fairness, and data dynamics.
  \item[—]
    It regards the EDIV problem as the outlier detection tasks, rather than following the mainstream problem-solving perspective.
\vspace{-1 em}
\end{itemize}    & High Accuracy                                                      \\ \cline{2-6} 
  & Li \emph{et al.} \ \ \cite{li2021cooperative} & 2021      &   \begin{itemize}[leftmargin=*]
  \item[—]
    It does not need TPA involved.
  \item[—]
    It can repair corrupted data replicas automatically.
\vspace{-1 em}
\end{itemize}                                                        &    \begin{itemize}[leftmargin=*]
  \item[—]
    It does not achieve batch verification and data dynamics.
  \item[—]
    It does not consider possible attacks due to the assumption of no byzantine edge nodes.
\vspace{-1 em}
\end{itemize}     & High Computation \& Communication Efficiency, High Accuracy                                                     \\ \cline{2-6}  & Duan \emph{et al.} \cite{duan2022edge} & 2022 &   \begin{itemize}[leftmargin=*]
  \item[—]
    Trust can be enhanced due to no TPA engaged. 
\vspace{-1 em}
\end{itemize}                                                        &     \begin{itemize}[leftmargin=*]
  \item[—]
    It does not take data dynamics and recovery into consideration. 
\vspace{-1 em}
\end{itemize} & Not Applicable \\ \cline{2-6} & Li \emph{et al.} \ \ \cite{li2022edgewatch} & 2022 &   \begin{itemize}[leftmargin=*]
  \item[—]
    It does not need TPA involvement.
  \item[—]
    It designs an incentive mechanism to motivate edge nodes well behaved.
  \item[—]
    It tailor-makes a consensus algorithm.
\vspace{-1 em}
\end{itemize}                                           & \begin{itemize}[leftmargin=*]
  \item[—]
    It offers a probabilistic integrity guarantee by sampling a proportion of data blocks.
  \item[—]
    It designs specifically for honest edge nodes, rather than from the DO/DU's perspective.
\vspace{-1 em}
\end{itemize}   & High Computation \& Communication Efficiency, High Accuracy, High Consensus Efficiency                 \\ \cline{2-6} & Zhao \emph{et al.} \cite{zhao2023data} & 2023 & \begin{itemize}[leftmargin=*]
  \item[—]
    It explores a smart inspection algorithm to pre-select unreliable data replicas for different data owners. 
  \item[—]
    It employs a smart contract to achieve fair verification.
\vspace{-1 em}
\end{itemize}                                               &   \begin{itemize}[leftmargin=*]
  \item[—]
    The unreliability metrics are not comprehensive.
  \item[—]
    It brings extra resource consumption brought by the usage of the smart contract. 
\vspace{-1 em}
\end{itemize} & High Computation \& Communication Efficiency                              \\ \hline \hline
\end{tabular}
\begin{tablenotes}
		\item 1: ``High'' indicates that this approach outperforms the corresponding comparison method(s) on average for this evaluation metric. Please refer to the corresponding papers for more experimental details.
\end{tablenotes}
\end{threeparttable}
}
\end{table*}

\subsection{Cooperative Audit}
\label{subsec:Cooperative Audit Papers}
As we mentioned before, private audit has fairness issues as it is the data owner/user that verifies integrity without the confirmation of edge nodes. Public audit is able to compensate for this limitation but hard to ensure that TPAs are totally trustworthy from the other participant's perspective, which may lead to potential security risks. To overcome this drawback, numerous works recently head to collaborative audit, in which edge nodes collaborate with each other to check EDI without TPA or even data owners/users' involvement. In that case, both the verification fairness issue and the TPA trusted issue can be eliminated naturally. We articulate the publications on cooperative audit in the following.

\par Alazeb \emph{et al.}~\cite{alazeb2019ensuring} (2019) focus on healthcare systems to detect malicious transactions by using a rule-based strategy. It is much more like an intrusion detection system, where edge nodes identify malicious behaviors that may corrupt data integrity according to specific corruption detection regulations. However, there are no experimental evaluations to validate their idea, which undermines the credibility of the proposed approach regarding follow-ups. Besides that, intrusion detection-based integrity verification is still a probabilistic method, and thus the accuracy needs to be studied further. Notably, this approach can not ensure fairness, as data owners are unable to participate in the verification process. Moreover, Yue \emph{et al.}~\cite{yue2020blockchain} (2020) restore blockchain and exploit a decentralized EDI sampling verification scheme in edge-cloud storage scenarios. Merkle tree with random challenging numbers is adopted and analyzed for system performance optimization. Additionally, they develop rational sampling strategies to address the problem of limited resources and high real-time requirements, making verification more effective. It keeps fairness because blockchain, as a third party, inspects integrity proofs for both the data owner and edge nodes. Nevertheless, it does not consider batch verification, data recovery, and dynamics issues.

\par In addition, John \emph{et al.}~\cite{john2020binary} (2020) explore several machine learning-based classifiers to check the integrity of electrocardiogram data. The feature vectors are derived from low complexity kurtosis and skewness~\cite{joanes1998comparing} based signal quality indices. The approach is more like~\cite{alazeb2019ensuring}, as both of them solve EDIV problems from the corruption detection perspective, rather than depending on interactive verification via \emph{Challenge-Response} mechanisms. The best part of it is that extensive experiments are conducted to evidence which machine learning model is the most suitable one for integrity corruption detection. It seems an ensemble of three neural networks using bagging with appropriate structures exhibits the best performance during testing for all the parameters considered with 99.47\% accuracy. From the utility point of view, this work saves lots of experiment simulation burden for researchers who would like to devote to this topic. However, due to the model-centric design, batch verification, fairness, data recovery, and dynamics are not investigated.

\par One step further, Li \emph{et al.}~\cite{li2021cooperative} (2021) propose the CooperEDI scheme to inspect EDI in a distributed manner. CooperEDI employs a distributed consensus mechanism to establish a self-management edge caching system. Edge nodes cooperatively ensure the integrity of cached replicas and repair corrupted ones. It rigorously considers data recovery problems but neglects computation efficiency and data dynamics. Although CooperEDI does not involve TPA, fairness is still hard to ensure since edge nodes may collude to generate incorrect verification results and the data owner can be only notified of corruption passively without knowing if the verification results are authentic and effective. Moreover, they fail to study how to secure against potential attacks such as byzantine attacks. Recently, they present EdgeWatch~\cite{li2022edgewatch} (2022), a collaborative EDIV framework, by leveraging blockchain. EdgeWatch collaborates with edge nodes to complete verifying a data replica cached on a certain edge node, and at the same time, the incentive mechanism is designed to motivate other edge nodes to join together into EDI inspection processes in a fast and honest way. The corresponding consensus algorithm is carefully designed to make edge nodes reach consensus. However, in practice, data owners who expect to check the integrity of outsourced data are always not edge nodes, so we argue that EdgeWatch has limited application scenarios. Very recently, Duan \emph{et al.}~\cite{duan2022edge} (2022) claim that it is impossible to avoid the collusion of edge nodes with malicious intruders. To solve it, they explore a blockchain-based verification protocol based on a distributed virtual machine agent that is an edge data integrity monitoring framework. In this way, trusted verification can be achieved without depending on a TPA. However, it does not support batch verification, fairness, data recovery, and dynamics. To further reduce verification overhead, Zhao \emph{et al.}~\cite{zhao2023data} (2023) make the first attempt to study the EDIV problem in the scenario with multiple data owners and multiple edge nodes. They introduce a smart inspection algorithm aimed at selecting unreliable data replicas before conducting the verification process for efficiency improvement. Nevertheless, it is imperative to comprehensively evaluate the reliability of data replicas, potentially by incorporating additional evaluation metrics.
\begin{figure*}[!tp]
\setlength{\abovecaptionskip}{0pt}%
\setlength{\belowcaptionskip}{0pt}%
	\centering
	\subfigure[Computation Efficiency \emph{vs.} $n$]{
		\begin{minipage}[b]{0.28\textwidth}
			\includegraphics[width=1\textwidth]{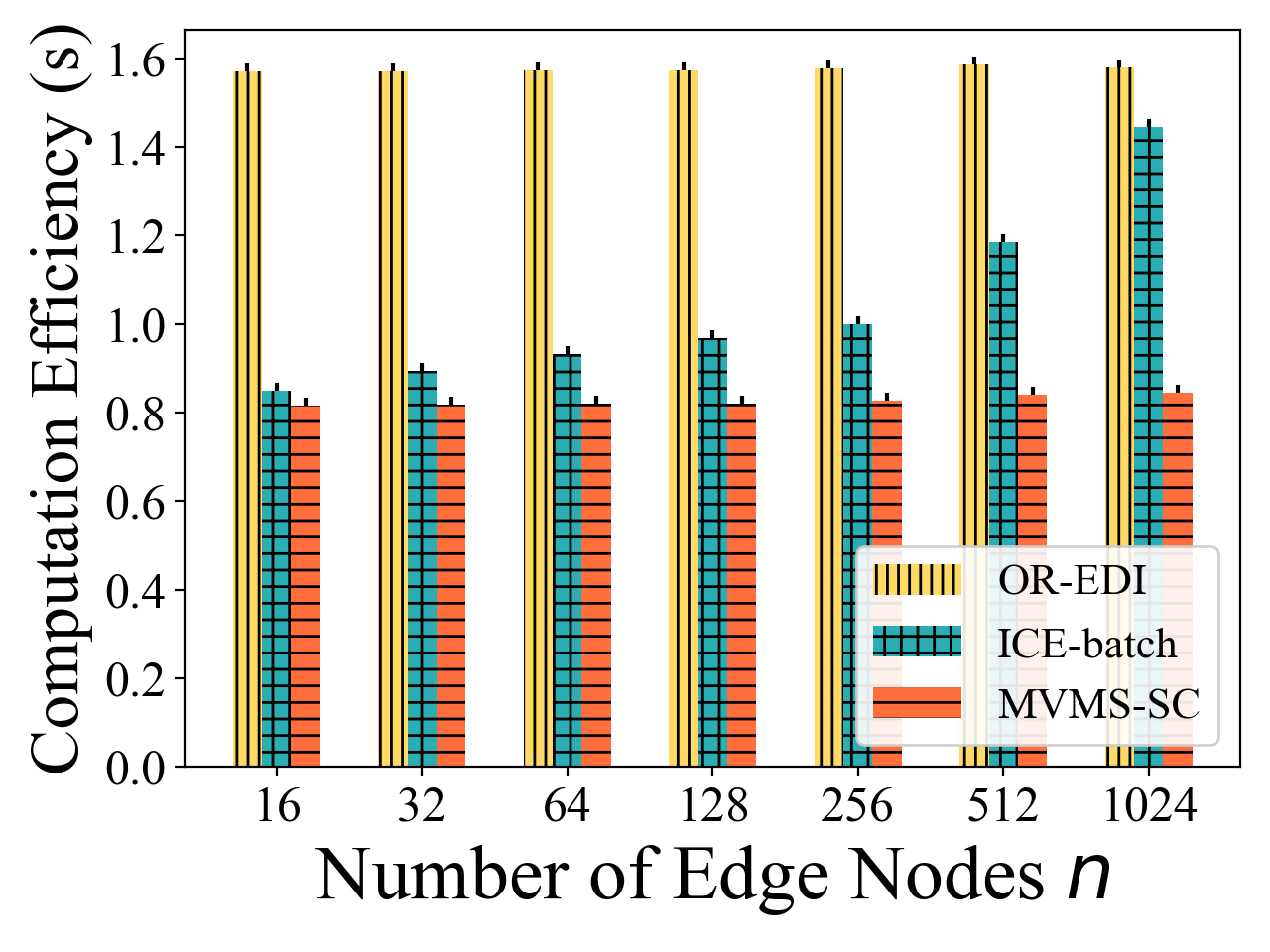}
		\end{minipage}
		\label{fig:computation efficiency_n}
	}\hspace{5mm}
    	\subfigure[Computation Efficiency \emph{vs.} $m$]{
    		\begin{minipage}[b]{0.28\textwidth}
   		 	\includegraphics[width=1\textwidth]{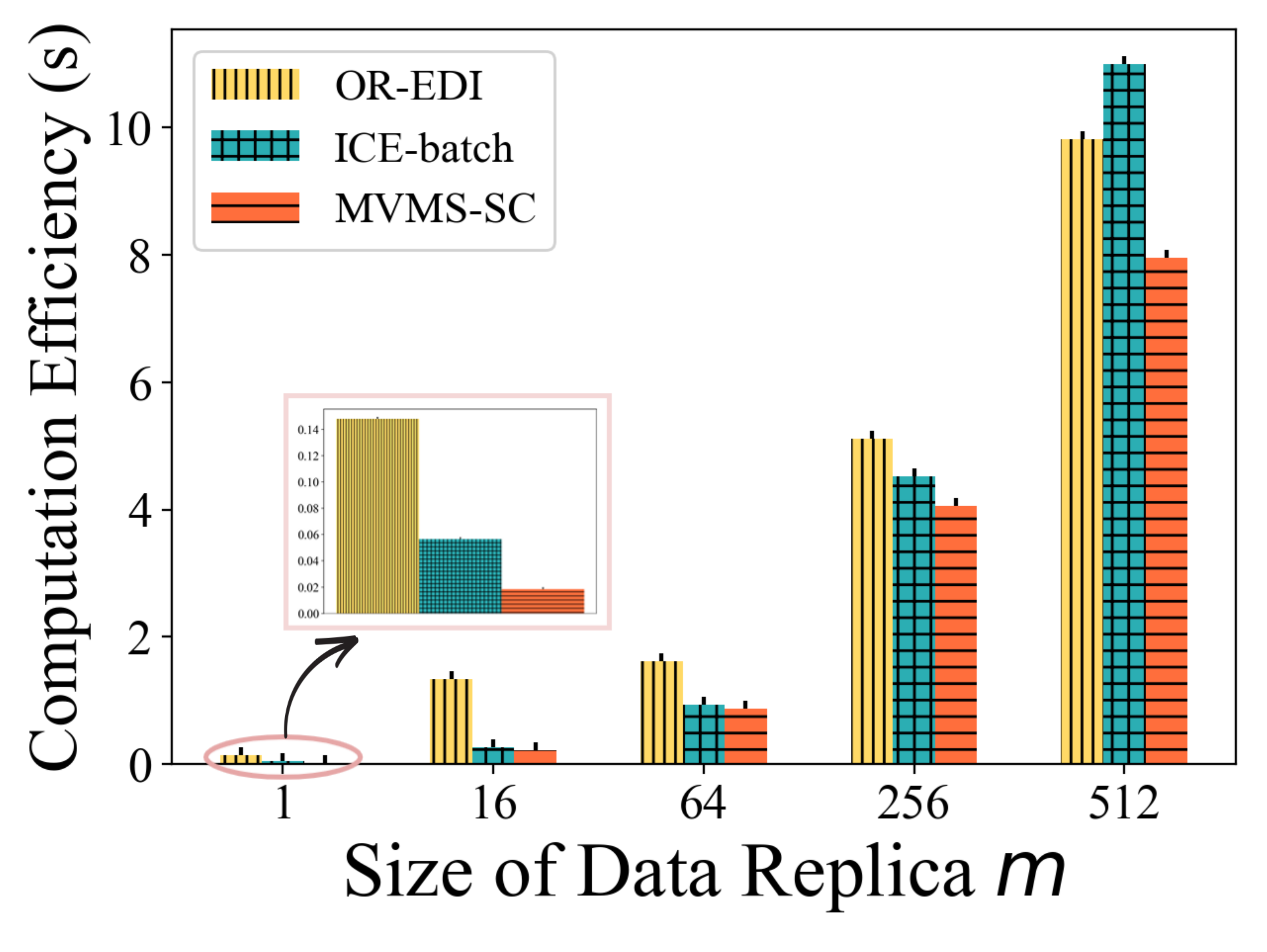}
    		\end{minipage}
		\label{fig:computation efficiency_m}
    	}\hspace{5mm}
    	\subfigure[Communication Efficiency \emph{vs.} $n$]{
    		\begin{minipage}[b]{0.28\textwidth}
   		 	\includegraphics[width=1\textwidth]{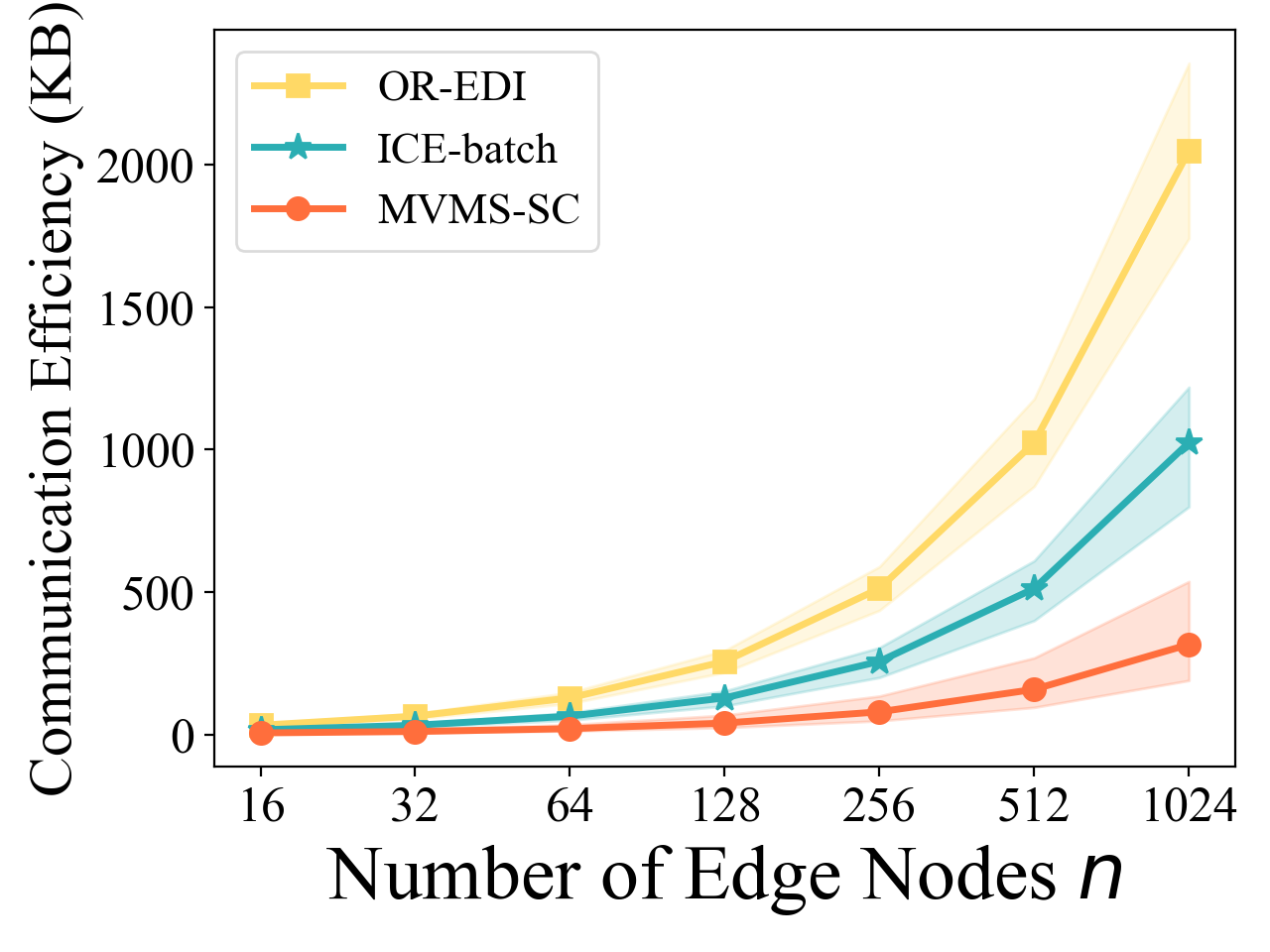}
    		\end{minipage}
		\label{fig:communication efficiency}
    	}
	\caption{Empirical analysis}
	\label{fig:Empirical analysis}
  \vspace{-1 em}
\end{figure*}
\subsection{Comparison of private, public, and cooperative audits}
\label{Comparison private, public, and cooperative audits}
For an intuitive comparison, we conduct empirical analysis for private, public, and cooperative audits. Notably, achieving a fair comparison among all existing methods is impractical due to diverse implementations, designs, and focuses even within the same category. As an illustration, we choose one state-of-the-art approach for each category, namely OR-EDI~\cite{li2023or} for private audit, ICE-batch~\cite{tong2022privacy} for public audit, and MVMS-EDI~\cite{zhao2023data} for cooperative audit. Such selection is based on the criteria that these three approaches are the latest ones and emphasize efficiency improvement.

\par Following the settings in~\cite{zhao2023data}, two parameters are adjusted to simulate different scenarios: (1) the number of edge nodes ($n$, same as the number of data replicas to be audited), which exponentially varies from 16 to 1024; and (2) the size of data replicas ($m$, MB as unit), ranging from 1MB to 512MB. Two key performance metrics are employed: \emph{Computation Efficiency} denoting the time taken by an approach to complete one round of EDIV, and \emph{Communication Efficiency} indicating the communication data size taken by an approach to complete one round of verification. 

\par Several observations come from the experimental results: (1) OR-EDI exhibits greater sensitivity to variations in the number of edge nodes $n$ compared to changes in the size of the data replica $m$; (2) ICE-batch demonstrates a more obvious growth trend in communication overhead when varying $n$ or $m$, compared to other approaches; and (3) MVMS-SC experiences significant efficiency benefits from the pre-selection mechanism.

\section{Open challenges and potential solutions}
\label{sec:Open challenges and potential solutions}
\begin{figure}[]
    \centering
    \includegraphics[width=1\linewidth]{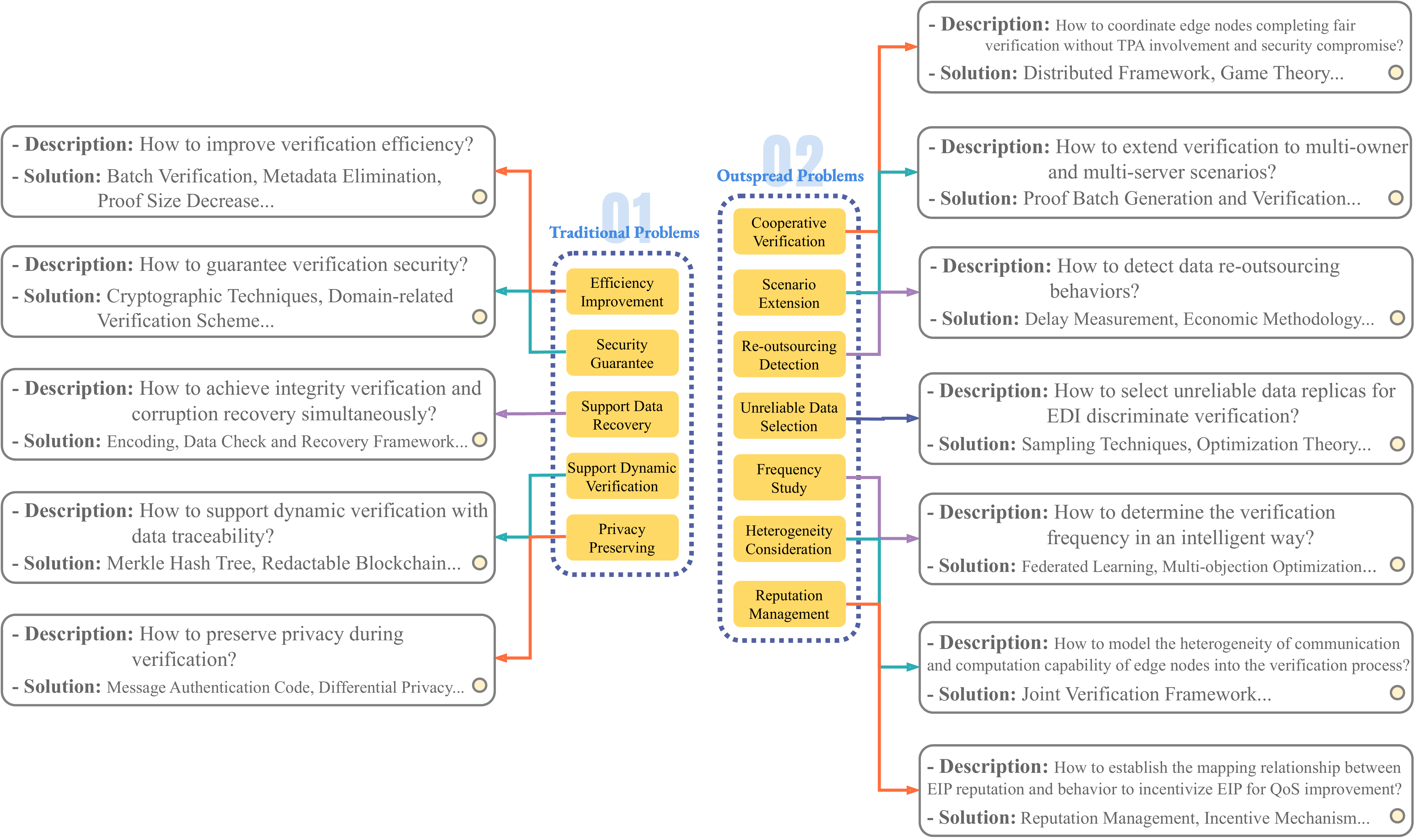}
    \caption{Future research directions and potential solutions for the edge data integrity verification problem}
    \label{fig:Future Research Direction for EDI}
\end{figure}

So far, we have predominantly witnessed positive advancements in the reviewed approaches designed to tackle data integrity issues within edge domains. While this encouraging progress may inspire further exploration by researchers, it is equally important to emphasize the practicality and adaptability of these existing technologies. In this section, we aim to outline and discuss acknowledged limitations in the literature we have reviewed, while shedding light on outspread challenges that have received less or even no attention, hoping to motivate future research in this domain. Open challenges are summarized in Fig.~\ref{fig:Future Research Direction for EDI}, encompassing both traditional and outspread problems.

\subsection{Traditional Problems}
\label{subsec:Traditional Problems}
In cloud computing environments, researchers commonly prioritize efficiency improvement, security assurance, data recovery, dynamic verification, and privacy preservation when developing integrity inspection solutions. These challenges also exist in edge computing scenarios, necessitating further study from diverse perspectives due to the uniqueness of edge environments.

\subsubsection{How to improve verification efficiency?}
\label{subsusec:Efficiency Improvement}
Integrity verification efficiency is a fundamental issue, directly influencing the practicability of EDIV solutions. Three aspects are specifically involved in terms of efficiency: computation, storage, and communication complexity. Batch verification is a common choice to reduce computation complexity~\cite{tong2019privacy,li2020auditing}. The proliferation of 5G networks has shifted the primary performance bottleneck of EDIV solutions from communication to computation. In this context, a more efficient batch verification scheme is required to satisfy stringent computing latency requirements. Regarding storage complexity, as edge nodes always cache additional metadata for future integrity verification, exploring ways to inspect EDI without retaining metadata presents a challenging yet promising avenue for storage reduction. One possible solution is to encourage data owners to store original data. In this way, both edge nodes and data owners could validate integrity without metadata engaged in the verification process. Lastly, communication efficiency in data integrity schemes involves three aspects: the transmission of original data to edge nodes, data owners/users initiating the verification procedure, and the transmission of integrity proofs in reverse. Such multi-round communications over long-distance backbone networks inevitably constrain overall efficiency improvements. Given the vulnerability of edge nodes, a high verification frequency is necessary to swiftly detect corruption, further imposing more workload on all participants. To tackle this challenge, privacy-preserving federated learning is a promising solution, wherein each edge node leverages the pre-trained corruption detection model for filtration, thereby circumventing the need for complex interactions. Besides, in the case of dynamic data, communication overhead also includes data updates. Another avenue for improving communication efficiency is to explore ways to reduce the size of integrity proofs, which is worth an in-depth study. 

\subsubsection{How to guarantee verification security?}
\label{subsusec:Security Guarantee}
With the integration of numerous new technologies into edge computing, the openness of the edge environment becomes both an advantage and a potential threat to its users. Data integrity schemes are susceptible to various attacks, as outlined in Table~\ref{tab:Review on existing internal attacks regarding the EDIV verification}. This vulnerability implies that certain existing EDIV solutions may not be entirely secure or reliable if attacks are successfully executed without timely detection. Additionally, the \emph{Challenge–Response} mechanism is vulnerable to data leakage attack if the proof generation method is not semantically secured. To this end, various blockchain-based approaches have been explored to mitigate potential data leakage. However, how to reduce the time consumption of edge nodes attending blockchain consensus is a challenging issue~\cite{miglani2022blockchain, miglani2020blockchain}. Moreover, a plethora of cryptographic techniques, such as multiProver zero-knowledge proof~\cite{grilo2019perfect} and Homomorphic Verifiable Tag~\cite{wang2013knowledge}, have been widely applied in existing works to achieve security objectives, while it is still far from satisfactory due to efficiency issues. Given the diverse security assumptions in different scenarios, it is reasonable to design a domain-related verification scheme with a comprehensive theoretical security analysis.

\subsubsection{How to achieve integrity verification and corruption recovery simultaneously?}
\label{subsusec:Support Data Recovery}
When data owners detect corruption in their outsourced data, they expect the damaged data to be fully recovered. Existing integrity check techniques achieving data recovery rely primarily on encoding methods, such as error-correcting code~\cite{chowdhury1994design} and network coding~\cite{katti2008xors}. However, these methods are effective only for low percentages of damage and entail a significant computational overhead. Therefore, it is critical to devise novel EDIV approaches with data recovery support. To tackle it, a collaborative data check and recovery framework could be employed. In brief, after identifying corruption, the edge nodes with corrupted data replicas could interact with nearby normal edge nodes to obtain the correct version. This collaborative approach enables the recovery of corrupted data with minimal communication overhead.

\subsubsection{How to support dynamic verification with data traceability?}
\label{subsusec:Support Dynamic Verification}
The outsourced data is dynamic by nature in EC environments, inherently subject to regular modification by data owners. Thus, offering support for dynamic operations on outsourced data is also critical for an auditing protocol. Edge nodes must update data strictly according to data owners' requirements to guarantee the correctness and timeliness of data. However, existing methods that support data dynamics, such as Merkle hash tree~\cite{niaz2015merkle} and index hash table~\cite{tian2015dynamic}, come with inevitable disadvantages. Specifically, Merkle hash tree requires substantial amounts of supplementary validation information to ensure the validity of data updates. Index hash table is effective only for modification, since the insertion and deletion operations disrupt the sequence structure of the original table, incurring additional costs. Furthermore, existing approaches often involve edge nodes caching only the most recent version of the data, with historical versions being deleted. In certain cases, data owners/users not only expect edge nodes to deliver the latest data block but also seek to access historical versions, which necessitates data traceability. Supporting data updates with data traceability is a valuable issue for EDIV in future development trends. An intuitive solution is to use redactable blockchain to trace historical versions and verification results in the long term.

\subsubsection{How to preserve data privacy during verification?}
\label{subsusec:Privacy Preserving}
Data privacy guarantee has always been a critical prerequisite in SLA for the development of edge caching services. A publicly auditable technique should safeguard privacy information from TPA, or TPA should be capable of auditing the owner's data without the risk of learning data content. Using message authentication codes (MACs) on the owner's data is one way to ensure privacy. During an audit, TPA challenges the integrity of randomly selected data blocks and their corresponding MACs. The edge node responds with a sequence of data blocks as well as the MACs, and TPA then checks the integrity of the data. However, this solution has the following drawbacks: (1) a linear sequence of data blocks is acknowledged to TPA, directly violating the privacy-preserving agreement between the data owner and TPA; (2) the communication and computation complexity scales linearly with the sampled block size; (3) audit cost can be very high if bandwidth between TPA and edge node is limited; and (4) it only supports static data files. Consequently, existing privacy-preserving EDIV systems are not perfectly feasible in reality. To this end, differential privacy~\cite{dwork2008differential} could be explored to handle this problem. If integrity proof is processed with differential private mechanisms, data privacy could be preserved but some noise is injected. This necessitates the development of a novel proof verification approach. With the prevalence of batch verification and a large number of interactions, the impact of noise can be mitigated, given that the noise typically conforms to the Laplace mechanism where the mean value is zero.

\subsection{Outspread Problems}
\label{subsec:Outspread Problems}
Aside from the concerns discussed in Section~\ref{subsec:Traditional Problems} which exist in both cloud and edge, the following issues are particularly prevalent in edge.

\subsubsection{How to coordinate edge nodes completing fair verification without TPA involvement and security compromise?}
\label{subsec:Cooperative Verification}
In recent years, the general trend of the development of EDIV is to enable edge nodes themselves to perform integrity verification, i.e., cooperative verification, due to the following two reasons: (1) releasing the assumption of trusted TPAs while keeping fairness; (2) making most of the communication overhead occurring in backhaul networks, instead of backbone networks to greatly reduces communication overhead. Until now, there have been lots of attempts focusing on this direction, like~\cite{yue2020blockchain}. Some of them, e.g.,~\cite{yue2020blockchain, li2022edgewatch}, try to adopt blockchain to replace TPAs for public audit, however, they are not aware of potential security risks brought by blockchain itself, e.g., outsourcing attacks \cite{benet2017proof} and byzantine attacks \cite{meena2013survey}. Besides, others, such as~\cite{li2021cooperative}, apply traditional distributed algorithms, e.g., Raft~\cite{ongaro2014search}, Paxos~\cite{lamport2001paxos}, to enable edge nodes to communicate with each other for EDI inspection tasks under the assumption of no byzantine edge nodes. Further investigation is required to address the challenge of releasing unrealistic assumptions while ensuring the honest behavior of edge nodes during cooperative verification. One potential approach to address this issue is leveraging game theory, which focuses on logical decision-making to ensure honest behavioral relations.

\subsubsection{How to extend verification to multi-owner and multi-server scenarios?}
\label{subsec:Scenario Extension}
Existing EDIV solutions only consider the case of EC domains with one data owner and multiple edge nodes. However, as we mentioned in Section~\ref{sec:Evaluation Criteria on Verification Approaches}, these solutions can not be directly extended to support multiple data owners with data integrity assurance due to verification efficiency and scenario heterogeneity issues. \emph{To date, only a limited amount of research~\cite{zhao2023data} has focused on the EDIV problem with multi-owner and multi-server.} One potential solution is to tailor-make an efficient proof batch generation method for edge nodes to improve proof generation efficiency and accordingly develop a proof batch verification approach for proof verification efficiency enhancement. With careful design, the approach could well fit in such complicated edge environments.

\subsubsection{How to detect data re-outsourcing behaviors?}
\label{subsec:Re-outsourcing Detection}
SLAs restrict the EIP's ability to preserve data in a certain geographic region at the granularity level of city, state, time zone, or political boundaries. Nevertheless, dishonest EIPs may relocate data owners' data to a third-party data centre, which usually has less computation and communication capacities, in breach of SLAs for saving storage space or enhancing profit. Undoubtedly, such malicious data re-outsourcing acts may conflict with the preferences of data owners and jeopardize their legitimate rights and interests, and worse than that, it might indirectly make data available to other governments, who can review it via search warrants or any other legal means, which invades sensitive data privacy, especially defense data. The common \emph{Challenge-Response} mechanism can not provide proof that data cached on untrusted edge nodes is not re-outsourced to other economical ones, especially in collusion network architectures. \emph{There is no relevant study on re-outsourcing detection in edge storage so far.} An intuitive solution is to simply measure the network delay of different distances. Clearly, it could not prevent untrusted edge nodes from re-outsourcing data to some other nearby yet cost-effective edge nodes. In this case, fast detection of intentional dishonesty or breach of re-outsourcing is critical for data owners/users. It may be addressed by economic methodologies, such as incomplete information dynamic game models.

\subsubsection{How to select unreliable data replicas for EDI discriminate verification?}
\label{subsec:Unreliable Data Replica Selection}
Existing EDIV solutions indiscriminately inspect all data replicas for data owners/users in each verification round~\cite{tong2019privacy, liu2020efficient, li2020auditing, 9459478, li2021inspecting}. In fact, it is not likely for the majority of data replicas to be corrupted by various faults or cyberattacks simultaneously~\cite{li2021cooperative}, and thus data owners/users are able to merely verify a part of unreliable data replicas in each round due to efficiency and cost-effectiveness concerns~\cite{chen2020bossa, zhang2018domain}. Indeed, some researchers~\cite{balamurugan2015multiple} have exploited a sampling-based method that supports inspecting partial data replicas by adopting a straightforward sampling technique, i.e., proportionally stratified sample~\cite{hirzel2002optimal}, in cloud computing environments. Obviously, this approach is neither reasonable nor tenable in real-world scenarios. To handle this issue, it is possible to adopt a dynamic selection process based on the optimization theory. The problem could be modeled as a constrained optimization problem by jointly considering the inherent property of data replicas and the performance of cache services (e.g., quality of service (QoS)~\cite{aurrecoechea1998survey}), and the (approximate) optimal solution could be derived by various optimization algorithms, e.g., simplex method~\cite{nelder1965simplex}, lagrangian multiplier method~\cite{everett1963generalized}, and genetic algorithm~\cite{whitley1994genetic}.

\subsubsection{How to determine the verification frequency in an intelligent way?}
\label{subsec:Frequency Study}
Substantial work on the EDIV problem, depends almost exclusively on the round-based \emph{Challenge-Response} mechanism that is invoked periodically at time intervals of a specified duration. In that case, verification frequency is one of the most fundamental problems for approach design and directly affects verification accuracy. \emph{Although extensive research is underway on the improvement of EDIV efficiency, none of them have rigorously considered the unlimited verification frequency property}, as shown in Table~\ref{tab:Summary and Qualitative Comparison of Existing Solutions}. They all focus on designing EDIV approaches from the one-round perspective. Nevertheless, studying verification frequency is significantly essential for approach practicability. More specifically, if EDI is inspected frequently through the \emph{Challenge-Response} mechanism, the computation and communication cost on both sides becomes extremely high. Instead, if setting a low frequency, corruption behaviors may not be found and corrected promptly, which may cause huge losses to data owners/users. Therefore, it is reasonable to work out an (approximate) optimal trade-off among verification frequency, verification accuracy, and resource consumption, regarding EDIV. The straightforward solution is that edge nodes collectively train a frequency selection model using federated learning, if there is enough training data with a large number of verification-related features e.g., inspection results and times. In practice, however, obtaining training datasets is often challenging, making this implementation inapplicable. In this case, multi-objection optimization algorithms, e.g., non-dominated sorting genetic algorithm $\uppercase\expandafter{\romannumeral2}$~\cite{deb2001controlled}, could be used to derive such a trade-off.

\subsubsection{How to model the heterogeneity of communication and computation capability of edge nodes into the verification process?}
\label{subsec:Joint Verification}
Although existing EDIV solutions provide high detection efficiency with relatively low overhead, their applicability is very limited as they totally rely on an implicit assumption, that is the edge nodes have the same computation and communication capability throughout the inspection execution. However, we have observed that not all the processes of edge nodes experience the same level of resource availability at exactly the same time in real-world cases. Thus, the edge node having adequate resources can adopt a complex yet accurate EDIV method, i.e., interaction verification through the conventional \emph{Challenge-Response} mechanism, but others may be incapable of employing it at the same speed. If the solution fully depends on this interaction for EDI assurance, it is hard to ensure the feasibility due to the decentralized distribution of edge nodes and heterogeneity of resource requirements. Recent work~\cite{zhao2023long} makes the first attempt to release this unrealistic assumption from a long-term perspective to select a subset of edge nodes for verification, but it is far from satisfactory due to neglecting the dynamics of the involved edge nodes and failing to incorporate an effective EDIV process. To this end, we plan to design a joint verification framework to handle this problem. Specifically, data replicas can be filtered first by edge nodes in a flexible way according to their available resources. By doing so, honest edge nodes would report identified corruption actively. However, it is not adequate to merely enable edge nodes do inspection tasks due to the existence of dishonest edge nodes. A joint interaction verification at a relatively low frequency is needed to detect corruption that is not trustily reported by malicious edge nodes, which unlocks the better performance of EDIV under a more practical assumption.

\subsubsection{How to establish the mapping relationship between EIP reputation and behavior to incentivize EIP for QoS improvement?}
\label{subsec:Reputation Management}
We have witnessed the emergence of a variety of EIPs over the last few years. AWS, Microsoft, Google, and IBM are a few examples of companies that apply the combined strength of edge nodes to provide data caching services. While collaboration among edge nodes enhances network capacity utilization, it introduces new system risks. Cryptographic techniques, often known as hard security measures, offer only partial solutions by ensuring data integrity. An edge node can be a valid member of a collaborative group and hence pass the standard cryptographic security tests. It might, however, purposefully report misleading measurement findings to acquire extra value at the expense of others. Soft security risks are the name given to this type of danger. In this context, trust and reputation management systems have the potential to combat such soft security concerns effectively, since there is a strong positive correlation between EIP reputation and cached data reliability~\cite{ma2018privacy}, i.e., prestigious EIPs are more likely to keep cached data replicas intact for securing competitive advantage. Technically, each EIP could be associated with a reputation value, which is updated based on the reliability of his cached data and further serves as one of the pathways to achieve EDI discriminate verification. Apart from designing an effective and satisfactory EIP reputation management system, developing an incentive mechanism to motivate EIPs to behave honestly during EDIV is also a pressing problem. We plan to develop a tailor-made incentive mechanism. If we adopt the credit as an example, then the credit is distributed to every EIP in proportion to its honest behaviors. The well-behaved EIP could gain more credits and accordingly have a higher possibility of being selected by data owners for data caching.

\section{Discussion}
\label{sec:Discussion}
From a holistic standpoint, the incorporation of emerging technologies, such as artificial intelligence~(AI), context-aware security, and micro-service, presents substantial potential for bolstering security within the EC framework.

\par \emph{AI}: It encompasses machine learning, deep learning, and federated learning. Utilizing advanced AI models enables anomaly detection, pattern recognition, and behavior analysis. These techniques can adapt to evolving threats and enhance the ability to detect and respond to security incidents in EC environments.

\par \emph{Context-aware Security}: Context-aware security mechanisms consider the specific context in which computing devices function. By taking into account contextual information, e.g., location, user behavior, and network conditions, security measures can be dynamically adjusted, providing a more adaptive and responsive security framework for EC.

\par \emph{Micro-services Architecture}: Micro-services advocate for a modular and decentralized approach to software development. Such architectural style improves EC security by isolating individual components, thereby minimizing the impact of potential breaches and facilitating easier updates and patches. Additionally, micro-services enable the implementation of tailored security measures for specific components in EC.

\par \emph{Secure Hardware Technology}: The advancement of secure hardware components, e.g., trusted platform modules and hardware security modules, contributes to building a more secure foundation for edge nodes. These hardware-based security solutions are effective in protecting sensitive data and cryptographic operations for EC systems.

\par \emph{Quantum-safe Cryptography}: Given the potential threat posed by quantum computing to traditional cryptographic algorithms, the adoption of quantum-safe or post-quantum cryptographic techniques becomes essential to ensure the long-term security of EC systems.

\par \emph{Threat Intelligence and Information Sharing}: Collaborative methods for threat intelligence and information sharing among edge nodes can strengthen the collective defense against emerging threats. Sharing insights regarding potential risks and vulnerabilities can enhance the overall security posture of the EC ecosystem.

\par In conclusion, strategically incorporating these technologies can strengthen the security landscape in EC. Beyond addressing the EDIV problem, they hold significant potential for tackling present and future challenges within the dynamic and evolving EC environment.

\section{Summary}
\label{sec:Conclusion}
Edge computing is an emerging research field that has inspired intense interest in edge security, especially in EDIV investigation. Given the scarcity of a detailed review on the EDIV-related topic in the open literature, this paper provided a thorough survey of various EDIV methodologies. We began by discussing the significance and uniqueness as well as the typical system models with corresponding key processes for the study of data integrity assurance in edge. Then, the comprehensive approach evaluation criteria were developed, followed by the discussion and comparison of recently advanced EDIV designs. Finally, we highlighted alarming challenges and presented future directions, along with further discussion. The EDIV problem is still in its infancy and will quickly mature in the future years for providing generic and versatile solutions. We expect that this survey will generate great attention in this emerging area and motivate more research efforts toward the satisfactory investigation of data integrity verification in edge domains.

\begin{acks}
This work was supported in part by the Australian Research Council under grant LP190100594, in part by the Taishan Scholars Program No.TSQN202211214 and tsqnz20230621, in part by the Shandong Excellent Young Scientists Fund Program (Overseas) No.2023HWYQ-113, and in part by Shandong Provincial Natural Science Foundation No.ZR202211150015.
\end{acks}

\bibliographystyle{unsrt}
\bibliography{main}

\appendix

\end{document}